\documentclass[lettersize,journal]{IEEEtran}
\usepackage{amsmath,amsfonts,amssymb,bm}
\usepackage{algorithmic}
\usepackage{algorithm}
\usepackage{array}

\usepackage{textcomp}
\usepackage{stfloats}
\usepackage{url}
\usepackage{verbatim}
\usepackage{graphicx}
\usepackage{cite}
\usepackage{color}
\usepackage{multirow}
\usepackage{subfigure}
\usepackage{todonotes}
\usepackage{booktabs} 
\usepackage[flushleft]{threeparttable}
\usepackage{balance}
\usepackage{subcaption}
\usepackage{framed}
\usepackage{xspace} 
\newcommand{\find}[1]{
    \setlength{\fboxrule}{1pt}
    \begin{center}\
        \noindent\fcolorbox{black}{gray!10}{
        \begin{minipage}{.92\linewidth}
            #1
        \end{minipage}
    }
    \end{center}
    \smallskip
}

\newcommand{\toolname}{IssueCourier\xspace}

\begin{document}

\title{IssueCourier: Multi-Relational Heterogeneous Temporal Graph Neural Network for Open-Source Issue Assignment}

\author{
\IEEEauthorblockN{
Chunying Zhou\IEEEauthorrefmark{1},
Xiaoyuan Xie\IEEEauthorrefmark{1}\IEEEauthorrefmark{3}\thanks{\IEEEauthorrefmark{3}Xiaoyuan Xie is the corresponding author.},
Gong Chen\IEEEauthorrefmark{1},
Peng He\IEEEauthorrefmark{2},
Bing Li\IEEEauthorrefmark{1}
}

\IEEEauthorblockA{\IEEEauthorrefmark{1}School of Computer Science, Wuhan University, China}

\IEEEauthorblockA{\IEEEauthorrefmark{2}School of Cyber Science and Technology, Hubei University, China}

\IEEEauthorblockA{zcy9838@whu.edu.cn, xxie@whu.edu.cn, chengongcg@whu.edu.cn, penghe@hubu.edu.cn, bingli@whu.edu.cn}
}

\maketitle

\begin{abstract}
Issue assignment plays a critical role in open-source software (OSS) maintenance, which involves recommending the most suitable developers to address the reported issues. Given the high volume of issue reports in large-scale projects, manually assigning issues is tedious and costly. 
Previous studies have proposed automated issue assignment approaches that primarily focus on modeling issue report textual information, developers’ expertise, or interactions between issues and developers based on historical issue-fixing records.
However, these approaches often suffer from performance limitations due to the presence of incorrect and missing labels in OSS datasets, as well as the long tail of developer contributions and the changes of developer activity as the project evolves.
To address these challenges, we propose \toolname, a novel Multi-Relational Heterogeneous Temporal Graph Neural Network approach for issue assignment. 
Specifically, we formalize five key relationships among issues, developers, and source code files to construct a heterogeneous graph. Then, we further adopt a temporal slicing technique that partitions the graph into a sequence of time-based subgraphs to learn stage-specific patterns. Furthermore, we provide a benchmark dataset with relabeled ground truth to address the problem of incorrect and missing labels in existing OSS datasets.
Finally, to evaluate the performance of \toolname, we conduct extensive experiments on our benchmark dataset. The results show that \toolname can improve over the best baseline up to 45.49\% in top-1 and 31.97\% in MRR.

\end{abstract}


\section{Introduction}
\IEEEPARstart{O}{pen-source} software projects mainly depend on issue tracking platforms such as Bugzilla, Jira, and GitHub \cite{aung2022multi} to manage and resolve software issues efficiently. 
When a new issue is created in a system, a project maintainer, often referred to as a \textit{triager}, is responsible for assigning the reported issues to appropriate contributors. The issue assignment process is far from trivial as it requires a deep understanding of the project, the issue and each member in the team, while also considering the past experience and current workload of each contributor. In traditional enterprise settings, this challenge is less pronounced, as developers typically have well-defined responsibilities, making it straightforward to assign issues to the designated maintainer of a specific module. In contrast, OSS development lacks such rigid role definitions, as contributors are often highly dynamic, frequently shifting between projects, and their areas of expertise may not be as clearly delineated as in proprietary software teams. These factors make manual issue assignment in OSS projects particularly complex, further motivating the need for automated solutions to improve efficiency and consistency in issue management.

In recent years, researchers have been actively exploring automated issue assignment approaches \cite{murphy2004automatic, mani2019deeptriage, he2021automatic, aung2022multi, jahanshahi2022s, dai2023graph, dai2024pcg, dong2024neighborhood, gousios2014exploratory, jahanshahi2023adptriage, yadav2024developer}. Existing approaches can be categorized into three types \cite{xie2021devrec}. The first is \textbf{content-based approaches} \cite{murphy2004automatic, mani2019deeptriage, he2021automatic, aung2022multi}, which use issue titles and descriptions as textual input and treat developers as classification labels.
The second is \textbf{collaborative filtering (CF)-based approaches} \cite{jahanshahi2022s, dai2023graph, dai2024pcg, dong2024neighborhood}, which model issue-developer relationships and often use graph-based structures to capture interactions and assignment patterns. The third is \textbf{expertise-based approaches} \cite{gousios2014exploratory, jahanshahi2023adptriage, yadav2024developer}, which rely on assessing developer expertise by constructing detailed developer profiles. 

Although previous works have proposed various algorithms to address different aspects of issue assignment, there are several limitations. 
First, we found that a significant number of issues in OSS repositories have missing or incorrect labels, which reduce the reliability of models trained on such datasets and hinder their effectiveness in real-world applications. Specifically, a substantial number of issues are closed, indicating that they have been resolved, but their \textit{assignee} fields, which serve as labels, remain empty. Moreover, in numerous cases, the developer officially assigned to fix an issue is not the one who actually resolves it.
Second, issue assignment models tend to favor experienced developers due to the long-tailed distribution of issue resolution by developers, and this tendency is further intensified by the presence of incorrect labels in the original dataset.
Third, traditional approaches overlook the fact that developer activity degrees change as the project evolves, which can cause models to incorrectly assign issues to developers who are no longer active. For example, a developer with an extensive history of issue resolution in the early stages but with little contributions in the recent stage. However, the model may still assign a significant number of issues to this developer because of their earlier contributions.

To address these limitations, we propose \toolname, a novel multi-relational heterogeneous temporal graph neural network approach for issue assignment. 
First, we extensively explore the complex relationships among issues, developers, and source code files to enrich the contextual information available for developers beyond their historical fix records, while also mitigating the challenges posed by the limited contributions of non-core developers. Furthermore, by segmenting the entire issue tracking timeline into distinct time slices, we aim to learn stage-specific patterns and adapt to temporal variations in developer activity by dynamically adjusting the influence of historical data. Besides, we provide a new benchmark dataset in which we have relabeled the issues, addressing the substantial amount of missing and incorrect labels.
In a nutshell, our contributions are as follows:

\begin{figure*}[!t]
\centering
\includegraphics[width=1.0\linewidth]{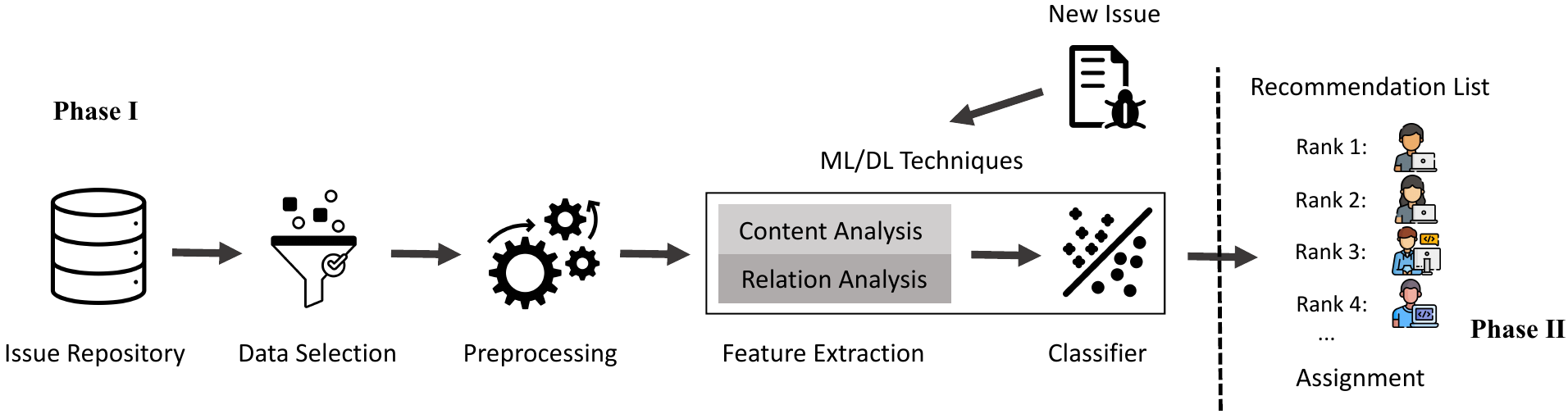}
\caption{General software issue assignment process.}
\label{issue-assignment-process}
\end{figure*}

(1) We provide a new benchmark dataset in which issues are relabeled to address the problems of incorrect and missing labels in the original dataset. Additionally, we conduct a manual assessment to verify the quality and reliability of the relabeled data.

(2) We model the complex relationships between issues, developers, and source code files, and propose \toolname for issue assignment based on the heterogeneous temporal graph, which explicitly encodes the above multiple relationships into the embedding learnings.

(3) We adopt a temporal slicing technique to transform the heterogeneous graph into a heterogeneous temporal graph, which helps the model effectively learn stage-specific patterns across different time stages.

(4) To evaluate the performance of \toolname, we conduct extensive experiments on the benchmark dataset. The experimental results demonstrate the superiority of \toolname over other state-of-the-art approaches. 

The rest of the paper is organized as follows. Section \ref{preliminaries} formalizes the issue assignment task and illustrates specific challenge examples that motivate our work. Section \ref{methodology} gives the overall framework and a detailed description of our approach. Section \ref{experimental setup} lists the experimental setup, and Section \ref{results and analysis} presents the experiment results and the analysis of research questions. Section \ref{discussion} discusses the impact of different labeling systems on the performance of existing methods. The threats to validity are discussed in Section \ref{threats to validity}. Section \ref{related work} gives an overview of the related work. Finally, we conclude this paper in Section \ref{conclusion}.

\section{Preliminaries}\label{preliminaries}

\subsection{Problem Description}

Issue assignment, also known as bug triage or bug assignment, is a crucial step in software maintenance, aiming to assign reported issues to appropriate developers for timely resolution. 
Once an issue report is submitted to the software issue repository, \textit{triager} first selects the similar issues that are already closed and fixed. From these issues, \textit{triager} makes a list of potential developers who can fix the issues based on their expertise \cite{kanwal2012bug}.
An issue report typically consists of several key components, such as the issue title, description, comments, reporter, and commenters, all of which provide valuable information for issue assignment.
Over recent decades, researchers have leveraged machine learning (ML) and deep learning (DL) techniques to analyze both content features (titles, descriptions, comments) and relational features (developer activity history, collaboration networks) for automated issue assignment, as shown in Fig. \ref{issue-assignment-process}. 

In this paper, we formalize the issue assignement process as follows.
Let $I = \{i_1,i_2,...,i_p\}$ denotes the set of all issues, $D = \{d_1,d_2,...,d_q\}$ denotes the set of all developers, where $p, q$ represent the number of issues and developers, respectively. 
In an OSS development platform, for an issue $i_i \in I$ recommends a set of suitable developers $\{d_1,d_2,...,d_n\} \subseteq D$, where $n$ represents the number of recommend developers. 

\subsection{Motivation}

\textbf{Motivation 1: Incorrect labels in OSS datasets lead to recommendations for developers who are not the true fixers of the issues.} 
\begin{figure}[!t]
\centering
\includegraphics[width=1.0\linewidth]{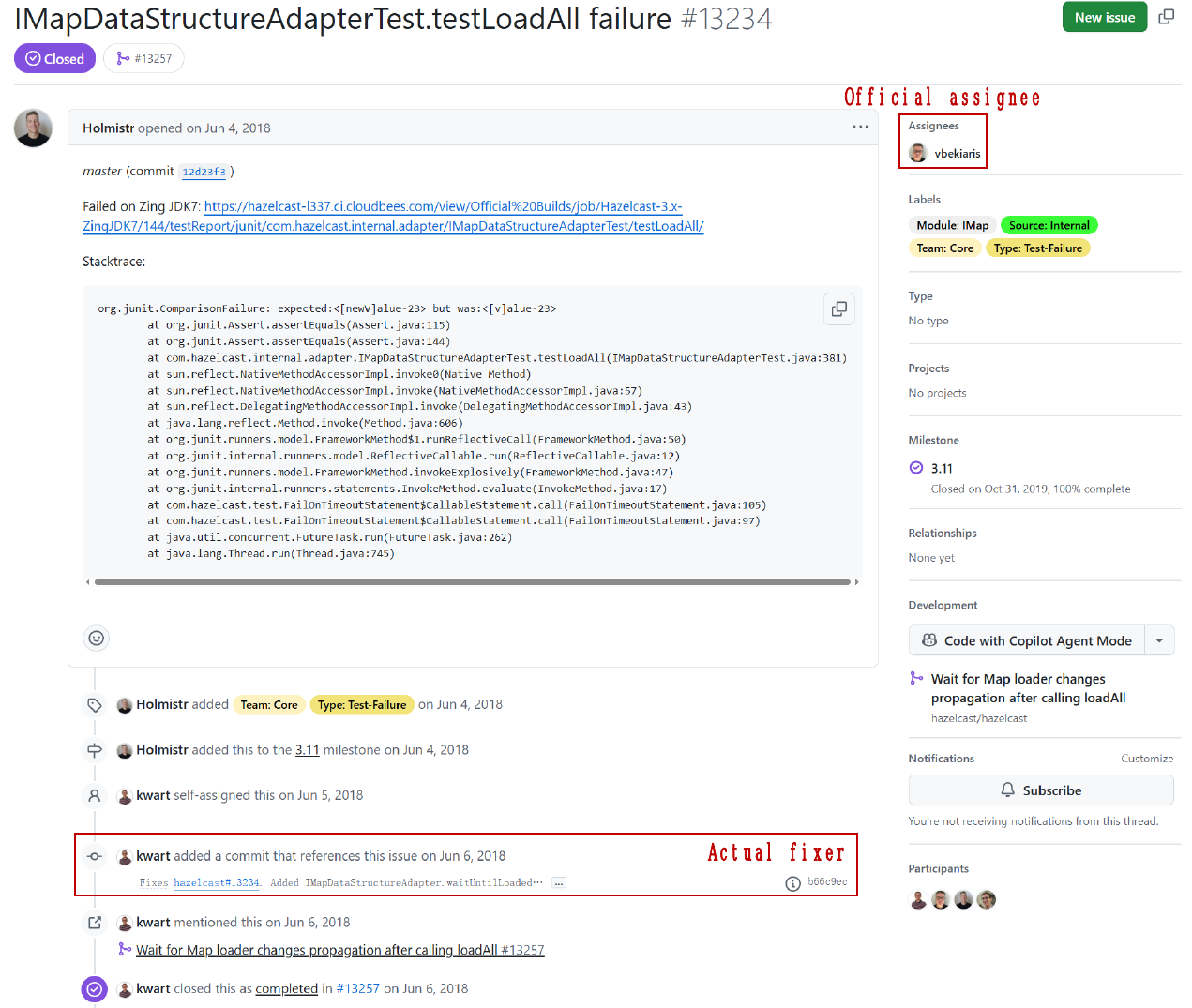}
\caption{The issue \#13234 is from the \textit{hazelcast\_hazelcast} project, which is officially assigned to \textit{vbekiaris}. The contributor \textit{kwart} is highlighted in red boxes to indicate his involvement in resolving the issue.}
\label{example_1}
\end{figure}
A key motivation we observed is the inconsistency between assigned developers and actual fixers. It is often the case that the developer who ultimately resolves an issue is not the one assigned to it. 
On the one hand, this inconsistency introduces incorrect labels, which can significantly affect model training by misleading the learning process. On the other hand, it can compromise evaluation, as models may appear to perform well by matching the original \textit{assignee} field, even though they fail to identify the actual fixer, thereby limiting their practical effectiveness.
We conducted a statistical analysis of \textit{hazelcast\_hazelcast} project on GitHub from 2012 to
2019 and found that among the 5,174 closed issues with an assigned developer, more than 900 of these
issues were actually resolved by other developers, which accounts for approximately 18\% of the total.
Fig. \ref{example_1} illustrates an issue from a Github project \textit{hazelcast\_hazelcast}\footnote{https://github.com/hazelcast/hazelcast/issues/13234}, where the assignee is identified as \textit{vbekiaris}. However, an examination of the issue's event log reveals that it was \textit{kwart} who actively committed the repair code during the resolution process, signifying their roles as the actual fixer of the issue.

\begin{figure*}
\centering
\begin{minipage}[t]{0.48\linewidth}
\centering
    \includegraphics[width=\linewidth]{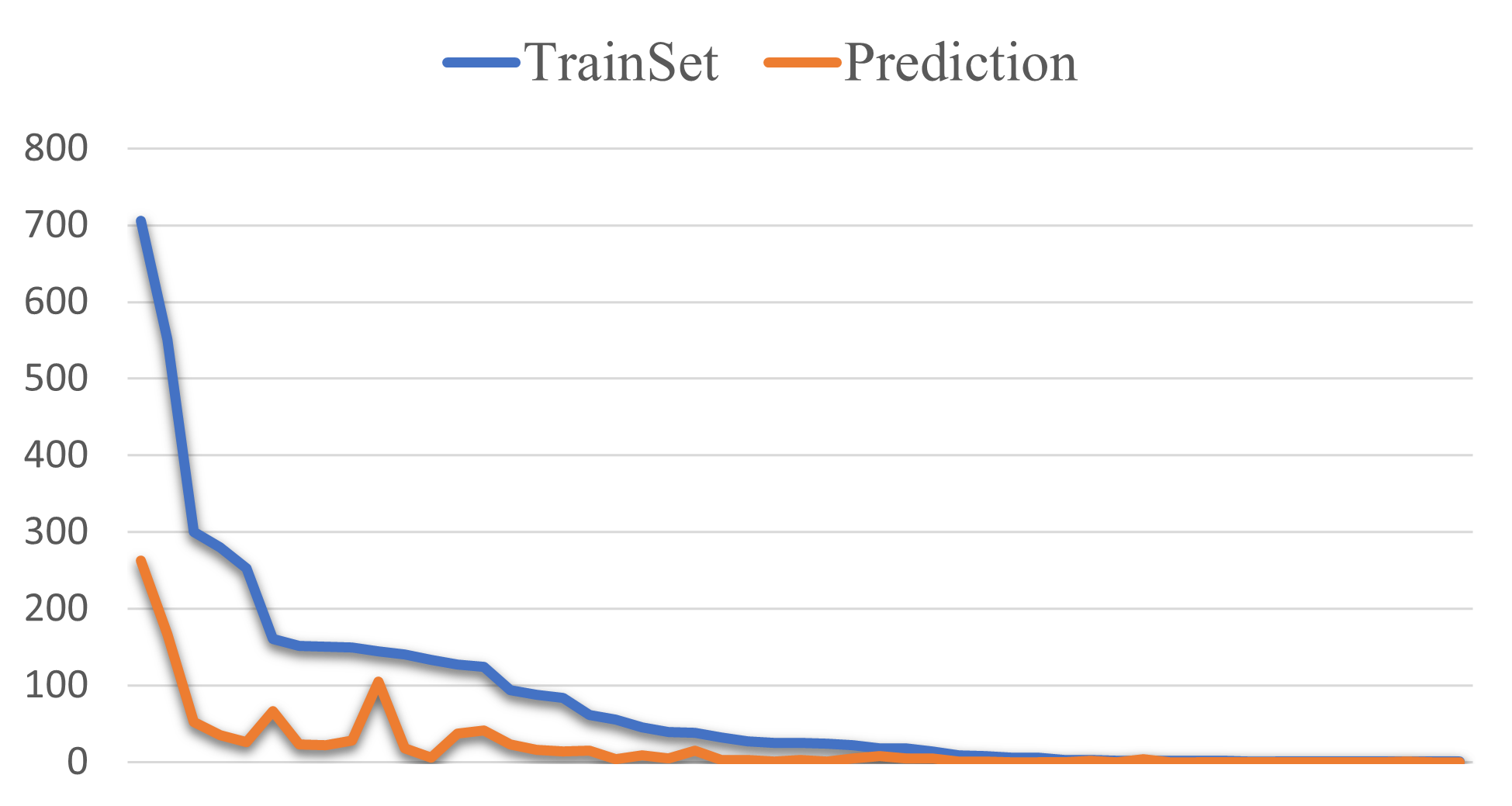}
    \caption*{(a) \textit{TrainSet} and \textit{Prediction} are based on the original labels.} 
\end{minipage}%
\begin{minipage}[t]{0.48\linewidth}
\centering
    \includegraphics[width=\linewidth]{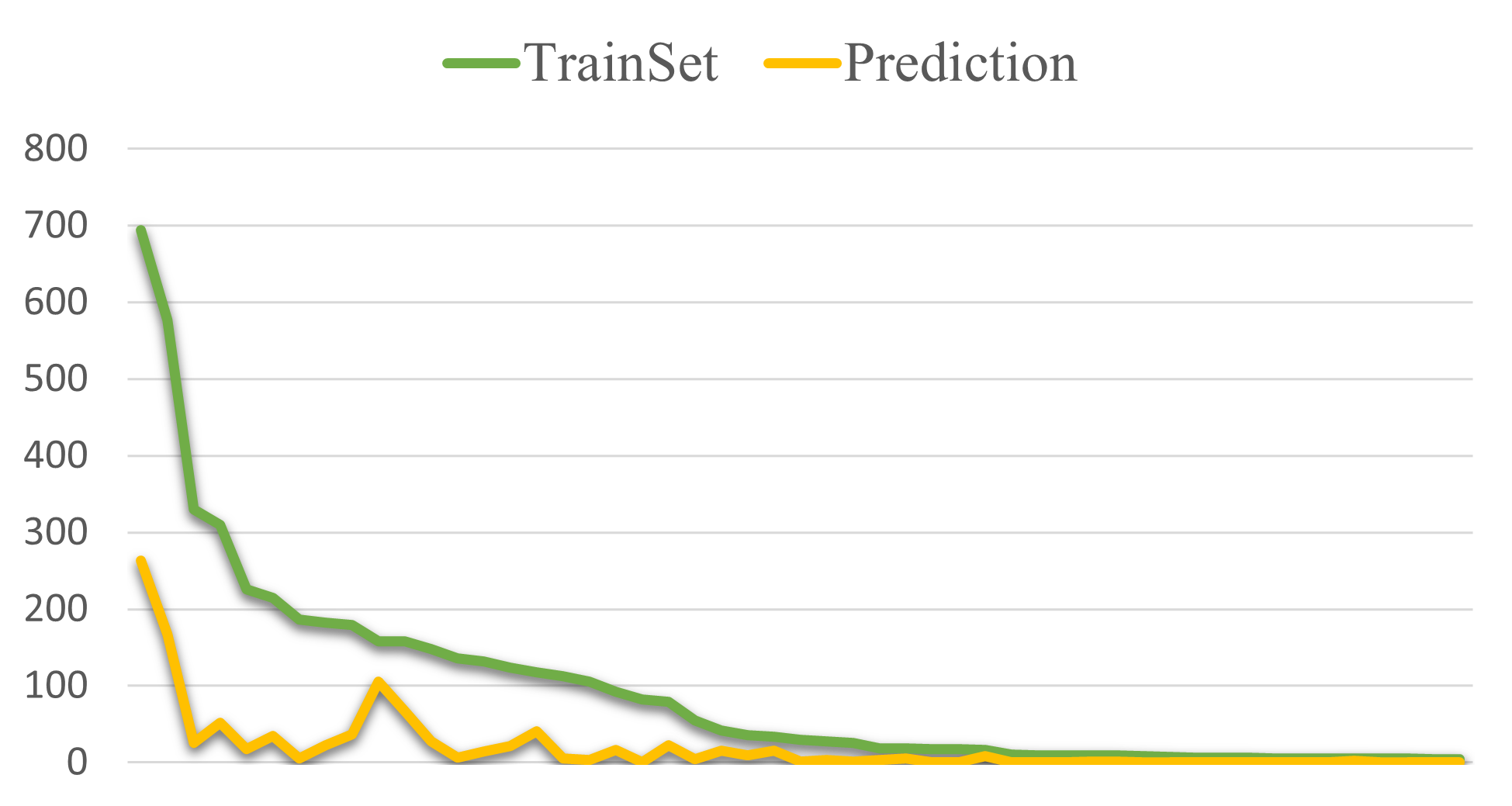}
    \caption*{(b) \textit{TrainSet} and \textit{Prediction} are based on the fixed labels.} 
\end{minipage}%
\caption{Distribution of the number of issues resolved by developers in the \textit{hazelcast\_hazelcast} project. In subfigures (a) and (b), \textit{TrainSet} represents the distribution of the number of issues resolved by developers in the training sets, based on the original and fixed labels, respectively. \textit{Prediction} denotes the prediction results of the Multi-triage approach \cite{aung2022multi} in the test set. Note that Multi-triage is trained on the original dataset, but predictions are evaluated using the original labels in (a) and the fixed labels in (b).}
\label{distribution}
\end{figure*}

\begin{figure*}[!t]
\centering
\includegraphics[width=0.8\linewidth]{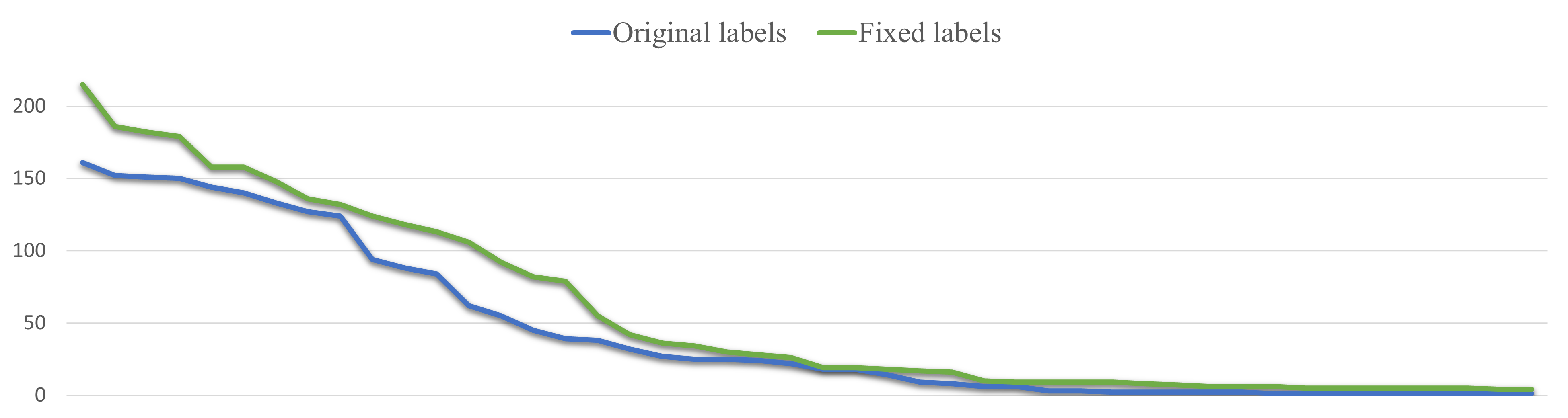}
\caption{The tailed distribution based on the original labels and fixed labels on the \textit{hazelcast\_hazelcast} project.}
\label{tailed}
\end{figure*}

\begin{table}[!t]
\caption{The activity degrees of the developers on \textit{hazelcast\_hazelcast} project.}
\label{preliminary-test}
\begin{tabular*}{\linewidth}{@{\extracolsep\fill}cccccc}
\toprule
\multirow{2}{*}{Developer ID} & \multicolumn{4}{c}{TrainSet} & \multirow{2}{*}{TestSet} \\\cmidrule{2-5}  
& Stage1 & Stage2 & Stage3 & Stage4 &  \\
\midrule

570462	& 5.83\% & 16.21\% & 14.88\% & 20.00\% & 19.40\% \\
4196298	& 0.08\% & 7.32\% & 15.53\% & 25.00\% & 10.93\% \\
1201712	& 16.19\% & 4.48\% & 3.78\% & 2.50\% & 1.91\% \\
2960360	& 12.24\% & 6.69\% & 3.86\% & 2.41\% & 4.10\% \\
105243	& 8.55\% & 4.17\% & 3.14\% & 2.59\% & 0.64\% \\
1741680	& 8.87\% & 7.32\% & 0.97\% & 0.17\% & 0.91\% \\
6005622	& 0.33\% & 2.83\% & 9.09\% & 2.84\% & 0.27\% \\
\textbf{\textcolor{blue}{3143425}}	& \textbf{\textcolor{blue}{3.70\%}} & \textbf{\textcolor{blue}{9.05\%}} & \textbf{\textcolor{blue}{1.77\%}} & \textbf{\textcolor{blue}{0.00\%}} & \textbf{\textcolor{blue}{0.00\%}} \\
158619	& 1.89\% & 3.54\% & 6.76\% & 2.33\% & 1.82\% \\
17194871 & 0.16\% & 0.24\% & 3.94\% & 8.97\% & 9.47\% \\
1212378	& 0.00\% & 0.24\% & 6.44\% & 6.47\% & 5.83\% \\ 
7543261	& 0.00\% & 0.94\% & 6.76\% & 4.48\% & 0.36\% \\ 
378108	& 5.51\% & 4.64\% & 0.80\% & 0.00\% & 0.27\% \\ 
234086	& 9.70\% & 0.79\% & 0.24\% & 0.09\% & 0.18\% \\ 
\textbf{\textcolor{blue}{1142801}}	& \textbf{\textcolor{blue}{7.81\%}} & \textbf{\textcolor{blue}{2.12\%}} & \textbf{\textcolor{blue}{0.16\%}} & \textbf{\textcolor{blue}{0.00\%}} & \textbf{\textcolor{blue}{0.00\%}} \\ 
3369598	& 0.00\% & 0.00\% & 4.34\% & 5.52\% & 5.92\% \\ 
7314547	& 2.79\% & 5.98\% & 0.16\% & 0.09\% & 0.00\% \\ 
823273	& 3.37\% & 4.01\% & 0.88\% & 0.26\% & 0.00\% \\ 
1047469	& 0.08\% & 1.02\% & 4.10\% & 2.33\% & 0.00\% \\ 
5988678	& 2.55\% & 1.10\% & 1.69\% & 1.38\% & 2.28\% \\ 
999867	& 0.00\% & 2.68\% & 2.49\% & 1.21\% & 0.73\% \\ 
2887123	& 0.58\% & 2.20\% & 1.37\% & 0.26\% & 0.00\% \\ 
5937333	& 1.31\% & 1.65\% & 0.24\% & 0.17\% & 0.09\% \\ 
2497208	& 1.31\% & 0.94\% & 0.56\% & 0.09\% & 0.00\% \\ 
719018	& 0.00\% & 0.08\% & 1.05\% & 1.72\% & 2.28\% \\ 
962661	& 0.00\% & 0.08\% & 0.08\% & 2.41\% & 9.47\% \\ 
\textbf{\textcolor{blue}{999984}}	& \textbf{\textcolor{blue}{0.66\%}} & \textbf{\textcolor{blue}{1.57\%}} & \textbf{\textcolor{blue}{0.00\%}} & \textbf{\textcolor{blue}{0.00\%}} & \textbf{\textcolor{blue}{0.00\%}} \\ 
510278	& 0.00\% & 1.18\% & 0.88\% & 0.00\% & 0.09\% \\ 
1912884	& 0.00\% & 0.16\% & 0.00\% & 1.47\% & 6.92\% \\ 
9109877	& 0.00\% & 0.00\% & 0.32\% & 1.29\% & 2.00\% \\ 
\textbf{\textcolor{blue}{3501420}}	& \textbf{\textcolor{blue}{0.66\%}} & \textbf{\textcolor{blue}{0.47\%}} & \textbf{\textcolor{blue}{0.32\%}} & \textbf{\textcolor{blue}{0.00\%}} & \textbf{\textcolor{blue}{0.00\%}} \\ 
\bottomrule
\end{tabular*}
\begin{tablenotes}
\footnotesize
\item[a] \textbf{Note:} \textit{TrainSet}, divided into \textit{Stage1} to \textit{Stage4}, and \textit{TestSet} represent the actual activity degrees of developers in each stage.
\end{tablenotes}
\end{table}

\textbf{Motivation 2: The long-tailed data distribution biases models toward recommending experienced developers.} 
In commonly used datasets, a small number of experienced developers resolve the majority of issues, resulting in a long-tailed distribution of developer contributions. This phenomenon is further aggravated by the incorrect labels mentioned in \textbf{Motivation 1}. Developers in the tail of this long-tailed distribution are typically non-core developers. Due to the sparse of their historical data, models tend to overlook them, as their exclusion does not significantly impact overall performance. However, this does not imply that non-core developers lack the capability to resolve issues. Instead, the limited assignment opportunities restrict their chances to gain experience and improve their skills.
We conducted a preliminary experiment using the Multi-triage \cite{aung2022multi} approach on the \textit{hazelcast\_hazelcast} project. Two key observations can be found from the results: Firstly, as shown in Fig. \ref{distribution}, a few core developers resolved over 50\% of the issues in the training set, leading the model to favor them while rarely recommending developers with fewer contributions. Even after we manually fixed the incorrect labels, the distribution of developer contributions still shows a long-tailed pattern. Secondly, as shown in Fig. \ref{tailed}, correcting the labels alleviates the long-tailed distribution phenomenon to some extent, with the contributions of non-core developers showing improvement compared to the original dataset with incorrect labels.

\textbf{Motivation 3: Ignoring the dynamic changes of developer activity as the project's evolution may cause misassignments.} 
In real-world OSS projects, developer participation and availability vary dynamically as the project evolves. Ideally, if a developer's recent activity degree is minimal, even if they accumulated many fixing records in earlier stages, models should reduce the influence of their early contributions during the prediction stage. However, existing models often overlook this temporal dynamic. Many of them treat historical data uniformly, which can result in assigning issues to developers who are inactive in the prediction stage.
Based on the \textit{hazelcast\_hazelcast} project, we quantified the degree of developer activity using their commit frequency and partitioned the dataset into five folds based on time, with the first four folds serving as training data (\textit{Stage 1} to \textit{Stage 4}), and the final fold serving as test set. It can be seen that the developers marked blue in Table \ref{preliminary-test} contributed during \textit{Stage 1} to \textit{Stage 3}, but their activity degree dropped to zero in the later training stage (\textit{Stage 4}). Despite this, a portion of issues during prediction may still be assigned to these developers, as the models relied on their contributions in earlier stages.

\subsection{Solution Strategies of IssueCourier}

To alleviate \textbf{Motivation 1}, we investigated the complex relationships between developers, issues, and source code files to enrich the profile of each developer. 
We recognize the existence of a beneficial relationship between developers who engage in issue discussions and those who report the issues. These individuals are likely to be members of the same team, collaboratively maintaining the functionality of a particular module, which accounts for their frequent co-occurrence in issue resolution processes. Therefore, we explore the \textit{report} and \textit{comment} relationships between issues and developers.
Furthermore, we also considered the relationships \textit{create} and \textit{remove} between developers and source code files, as developers more proficient in certain code modules may be better suited to handle issues related to those modules. 
The \textit{similar} relationship between issues and source code files can also serve as a useful auxiliary, as issue descriptions may sometimes reference corresponding code modules.
These relationships enable us to gather more information about non-core developers, allowing for appropriate issue recommendations even when they have little or no historical issue-fixing records. 
To tackle \textbf{Motivation 2}, we construct a heterogeneous temporal graph based on the above relationships and employ a temporal slicing technique to segment the data into multiple time periods. By aggregating information from adjacent time slices, this technique captures stage-specific patterns across different time stages and effectively balances the influence of developers' early contributions and recent activities, helping achieve more stable and reliable issue assignment.
Last but not least, to address \textbf{Motivation 3}, we provide a new benchmark dataset in which issues are relabeled by tracing issue events and identifying the developers who submitted commits related to the issue. This relabeling process aims to mitigate the mislabeling problem in OSS, ensuring that models are trained and evaluated based on the actual issue fixers rather than the originally assigned developers. 
To ensure the accuracy of this benchmark dataset, we have conducted a manual assessment to verify the correctness of the relabeled data.

\section{Methodology}\label{methodology}

\begin{figure*}[!t]
\centering
\includegraphics[width=1.0\linewidth]{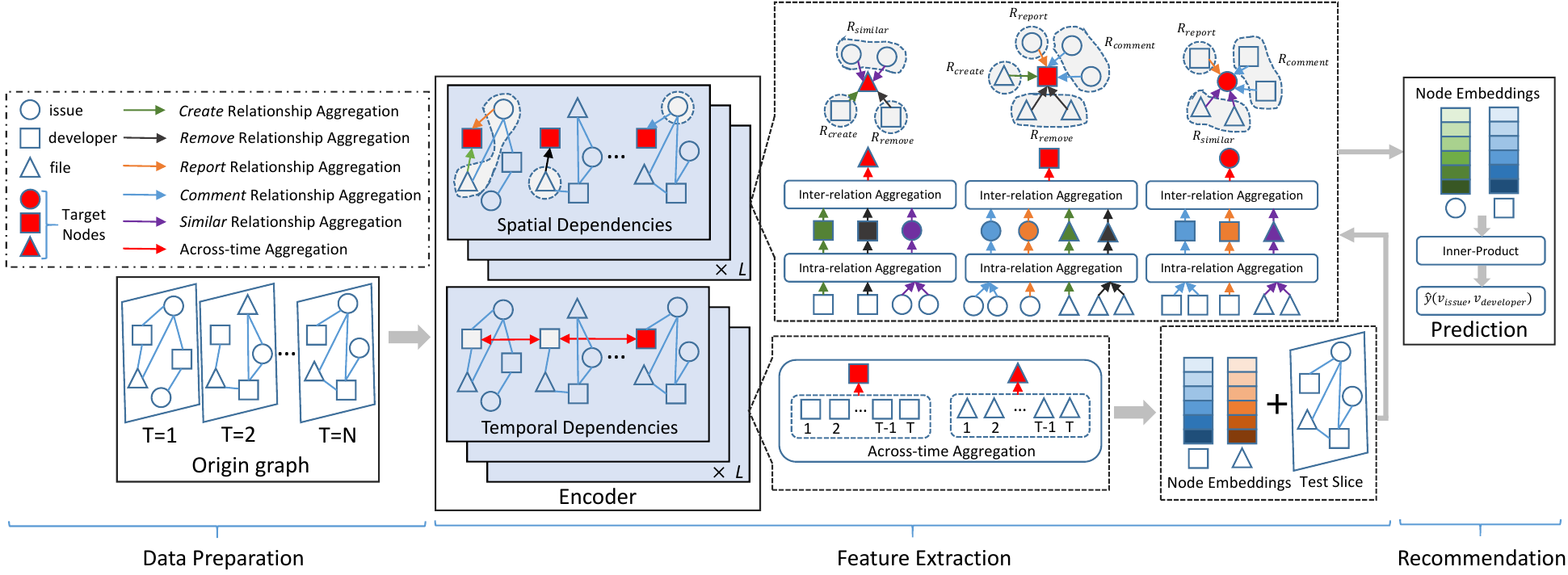}
\caption{The overall framework of \toolname.}
\label{framework}
\end{figure*}

% overview
This section presents the framework of our approach \toolname, which involves three steps shown in Fig. \ref{framework}: data preparation, feature extraction, and recommendation. In the data preparation phase, we relabel each issue by tracing the issue events and identifying the actual fixes. Based on this, we extract the \textit{report} and \textit{comment} relationships between developers and issues, the \textit{create} and \textit{remove} relationships between developers and source code files, and the \textit{similar} relationship between issues and source code files, thereby constructing a heterogeneous temporal graph (HTG) \cite{fan2022heterogeneous}. In the feature extraction phase, \toolname learns the information within each time slice and captures the evolution of information across adjacent time slices, allowing the model to effectively encode both the local structural context (spatial dependencies) and the temporal dynamics (temporal dependencies). Finally, in the recommendation phase, \toolname calculates the cosine similarity between the learned representations of issues and developers, and recommends the top N developers for a given issue.

\subsection{Data Preparation}

The datasets used in our paper are extracted from a contribution database \cite{diamantopoulos2020employing}, which includes the individual contributions of more than 60,000 contributors across 3,000 GitHub repositories. All data is hosted in a MongoDB instance and is organized in collections that include information on issues, issue comments, issue events, commits, repository statistics, and developer activity metrics, comprising more than 800,000 issues, 1,800,000 comments, 2,300,000 events, and 3,900,000 commits.

\subsubsection{Data selection and Relabeling}

We sorted the 3,000 GitHub repositories based on the number of closed issues in descending order. However, it was observed that numerous projects in the database had missing information from the data extraction process. Specifically, in the MongoDB database, each issue in the \textit{issues} table is supposed to contain comment URLs and event URLs that can be found in the \textit{issue comments} table and \textit{issue events} table. However, there were instances where the comment URLs and the event URLs in the \textit{issues} table could not be matched with records in the \textit{issue comments} table and \textit{issue events} table. Consequently, we took the \textit{issues} table as the benchmark, extracted the corresponding comment URLs and event URLs for each issue, and recrawled the information for these URLs. Additionally, the \textit{commits} table also suffered from missing information, so we saved all commit URLs that appeared in the \textit{issue events} table and recrawled the data for these URLs. 

Ultimately, we selected five projects with the highest number of issues as our datasets and further subjected them to a relabeling process. We relabeled the issues by tracing the issue events and identifying the developers who submitted commits related to the issue. If multiple developers submitted commits that modified the code relevant to the issue, we labeled all of them as fixers. We believe that these developers possess the necessary expertise to resolve the issue, as their contributions indicate familiarity with the affected components of the code. Consequently, we labeled all developers who made commits before the issue was closed as fixers. If no valid commit could be traced, we labeled the developer who closed the issue as the fixer, provided their number of commits exceeded the issues they closed, ensuring they were actively involved in development rather than solely handling operations tasks \cite{matsoukas2020towards}.

To evaluate the quality of our relabeled datasets, we employ a dual assessment mechanism \cite{bryman2016social} whereby two independent evaluators (the first author of this paper and a graduate student serve as evaluators) independently label the issues following the above relabeling process. Given the inability to manually label 58,306 issues, we use the sampling method \cite{Eisenhower2016ElementsOS} to select the minimum number (MIN) issues which could ensure that the estimated population is within a certain confidence interval at a certain confidence level. Based on this method, we randomly sample 383 issues for manual evaluation. Subsequently, we use Cohen’s kappa coefficient \cite{chmura2002kappa} to measure the agreement of the assessment results between the evaluators. The Kappa score of the assessment results is 0.89, indicating a high degree of consistency between the evaluators, thereby demonstrating the accuracy and reliability of our relabeled datasets.

\subsubsection{Relation Extraction}

\begin{figure*}[!t]
\centering
\includegraphics[width=0.8\linewidth]{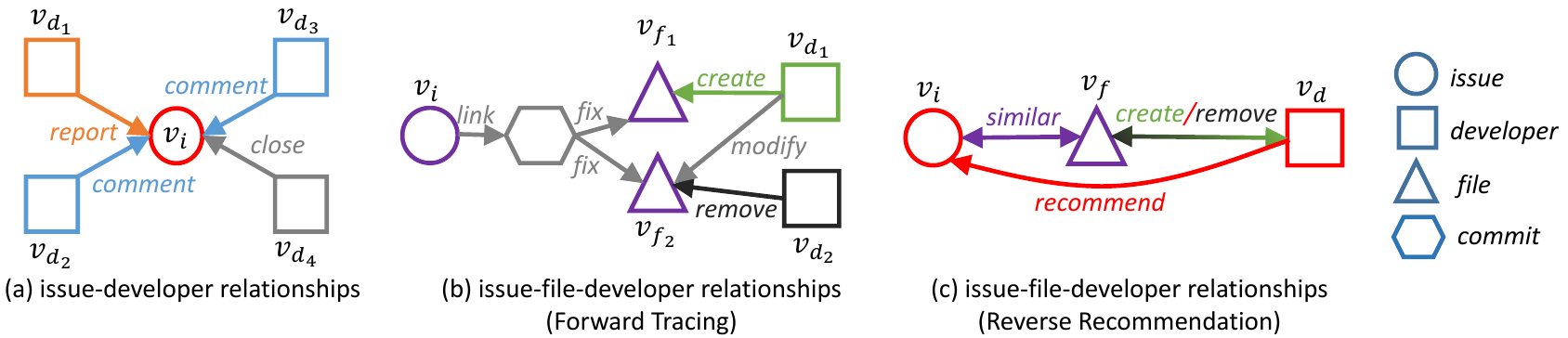}
\caption{Multi-relationships among issues ($v_i$), developers ($v_d$) and source code files ($v_f$).}
\label{relations}
\end{figure*}

The relationships among issues, developers, and source code files are inherently complex, with the direct relationships between issues and developers of importance in the context of the automatic issue assignment task.
Intuitively, a developer is more likely to handle an issue when there is a relevance between the developer and the issue. 
As illustrated in Fig.\ref{relations} (a), developers can play multiple roles in their interactions with an issue. For instance, a developer may be the individual who initially discovers and reports the issue, with a substantial likelihood of subsequently taking personal responsibility for resolving it. Furthermore, developers engaged in discussions about a specific issue are often part of the same team or collaborate on maintaining particular modules of the project. 
These relationships can be used to identify potential developers to resolve issues.
Here, we focus on extracting the \textit{report} and \textit{comment} relationships between issues and developers. Although there are other complex relationships between issues and developers, such as the \textit{close} relationship, which only exists for resolved issues. For a new issue, the easiest interactions to capture are limited to the \textit{report}, which initiates the lifecycle of the issue, and the subsequent \textit{comment}, which facilitate ongoing discussions and collaborative efforts toward resolution.

In addition to the direct interaction relationships between issues and developers, we also utilize the relationships between issues and source code files, as well as between developers and source code files. Specifically, as exemplified in Fig.\ref{relations} (b), by tracing the resolution process of an issue, we can retrieve the corresponding commit information, which contains details about the changed source code files. This constitutes a forward tracing progression in the issue resolution process, linking an issue to a commit, and subsequently to locate the code files.
In the case of a new issue, where no resolving commit has yet been submitted, we establish the auxiliary \textit{similar} relationship directly by calculating the textual similarity between the issue description and the content of these code files.
Furthermore, by analyzing all historical commit information, we can determine which modules developers frequently fix or are familiar with, such as the code files they create, modify, or remove. By utilizing this comprehensive understanding of developer-file relationships, we can indirectly recommend developers to issues related to their familiar code files (as shown in Fig.\ref{relations} (c)). 
Here, we selectively extract the \textit{create} and \textit{remove} relationships between developers and source code files.
Given the fact that any developer has the potential to fix any code file, and considering the high frequency of modifications to source code files, we focus solely on the \textit{create} and \textit{remove} relationships. Developers who have the authority to create or remove a source code file are likely to have a higher level of familiarity with it, as these actions require a deeper understanding and permission.

After extracting these relationships, we proceed to construct a HTG. We divide the timeline into ten periods based on the number of issues, rather than equal time intervals. Using equal intervals would result in large differences in the number of issues across time slices. Some slices would contain only a few dozen issues, while others could have more than a thousand. This imbalance could lead to overfitting in certain time slices. To avoid this problem, we segment the data so that each time slice contains a roughly equal number of issues. Within each of these time slices, we incorporate the developers associated with these issues, all source code files within that time period, and the relationships among them, collectively forming the heterogeneous graph for that specific time slice.

\subsection{Feature Extraction}

After constructing the HTG, we employ multiple aggregation modules to integrate the spatial and temporal dependencies information of each node, thereby extracting rich feature representations. A HTG is a sequence of snapshots, which is defined as $G=\{G_t\}_{t=1}^T$, where $T$ is the length of the time slice sequence. $G_t=\{V_t,E_t\}$ is a heterogeneous graph at time period $t$, where $V_t$ and $E_t$ are the temporal nodes and edges in $G_t$. Each node $v \in V_t$ has a type $\phi(v)$, and each edge $e \in E_t$ has a type $\psi(e)$. The node type set and edge type set are denoted by $A=\{\phi(v): \forall v \in V_t\}$ and $R=\{\psi(e): \forall e \in E_t\}$, with the constraint of $|A| + |R| > 2$. In this context, the node types include the issue nodes, the developer nodes and the file nodes, while the edge types include the \textit{report} and \textit{comment} relationships between the issue nodes and the developer nodes, the \textit{create} and \textit{remove} relationships between the developer nodes and the file nodes, and the \textit{similar} relationship between the issue nodes and the file nodes. 
Thus, $A = \{v_{i}, v_{d}, v_{f}\}$, and $R =$ $\{R_{report}, R_{comment}, R_{create}, R_{remove}, R_{similar}\}$.

\subsubsection{Spatial Dependencies}
In a time slice graph $G_t$, each type of node may have its own feature space. 
We adopt a type-specific projection on each node to map its distinct feature vector into a same feature space, which includes intra-relation and inter-relation aggregations.

\textbf{Intra-relation Aggregation.}
As illustrated in Fig.\ref{framework}, consider a developer-type target node $v_d$, which is connected to two types of neighboring nodes (i.e., issue type and file type), as well as four types of relationships. The relationships $R_{report}$ and $R_{comment}$ link $v_d$ to the issue nodes $v_{i_1}$ and $\{v_{i_2}, v_{i_3}\}$, respectively. Similarly, $R_{create}$ and $R_{remove}$ link to the file nodes $v_{f_1}$ and $\{v_{f_2}, v_{f_3}\}$, respectively. For each relationship type, intra-relation aggregation is first applied when there are multiple neighboring nodes with the same type and relationship. After aggregation, the updated representations of the four nodes are $v_{i_1}^{'}, v_{i_{\{2,3\}}}^{'}, v_{f_1}^{'}, v_{f_{\{2,3\}}}^{'}$.

Formally, given a target node $v$ at time period $t$ and a relation type $r \in R$, the intra-relation aggregation can be described as:
\begin{equation}
\label{eq1}
\textbf{h}_{v,r}^{t,l} = AGG_{intra}(\{\textbf{h}_{u,r}^{t,l-1}|u \in N_r^t(v)\}; \Theta_{intra})
\end{equation}
where $N_r^t(v)$ represents relation-based neighbors of node $v$ at time period $t$, $\textbf{h}_{u,r}^{t,l-1}$ is the embedding of node $u$ at time period $t$ in layer $l-1$, $\textbf{h}_{v,r}^{t,l}$ denotes the embedding of the relation $r$ with respect to the node $v$, and $\Theta_{intra}$ is the trainable parameters and is non-shareable for relation type, time period and aggregation layer. $AGG_{intra}(\cdot)$ is the aggregator function in intra-relation aggregation module.

\textbf{Inter-relation Aggregation.}
Through the process of intra-relation aggregation, the target node would gather multiple relation embeddings. 
Based on this, the inter-relation aggregation module aims to learn a spatial embedding for the target node summarizing the information from its spatial neighbors across all relation types. 
For instance, the developer-type target node $v_d$ aggregates information from the four types of neighboring nodes (i.e., $\{v_{i_1}^{'}, v_{i_{\{2,3\}}}^{'}, v_{f_1}^{'}, v_{f_{\{2,3\}}}^{'}\}$), resulting in the updated node representation $v_d^{'}$.

Formally, this process can be represented as:
\begin{equation}
\label{eq2}
\textbf{h}_{v,R}^{t,l} = AGG_{inter}(\{\textbf{h}_{v,r}^{t,l}|r \in R(v)\}; \Theta_{inter})
\end{equation}
where $R(v)$ denotes the set of relations connected to node $v$, $\textbf{h}_{v,r}^{t,l}$ is the embedding of relation $r$ with respect to node $v$ from intra-relation aggregation module, $\textbf{h}_{v,R}^{t,l}$ represents the spatial embedding of node $v$ that will be learned in this inter-relation aggregation module, and $\Theta_{inter}$ is the trainable parameters and is non-shareable for time period and aggregation layer. $AGG_{inter}(\cdot)$ is the aggregator function in inter-relation aggregation module.

\subsubsection{Temporal Dependencies}
While the spatial dependencies module aggregates information from a target node’s neighbors within the same time slice, the temporal dependencies module focuses on capturing across-time information by aggregating features from the node’s temporal neighbors across different time slices.

\textbf{Across-time Aggregation.}
We define the temporal neighbors as the same nodes in different time slices (including itself). This module takes the spatial embeddings of the target node’s temporal neighbors as input and outputs a spatial-temporal embedding for this target node. For example, the updated developer-type target node $v_d^{'}$ aggregates the temporal neighbors $\{v_{d_1}^{'}, v_{d_2}^{'}, ..., v_{d_T}^{'}\}$, resulting in the updated spatial-temporal embedding $v_{d}^{''}$.
It is noteworthy that within this module, only the developer-type and file-type nodes are capable of aggregating their temporal neighbors. This restriction arises because issues from different time slices do not overlap, ensuring that each issue within a given time slice is unique. Consequently, during this module, only the developer-type and file-type nodes can obtain their temporal dependency features, which can then be utilized in the subsequent time slice.
This process is formalized as follow:
\begin{equation}
\label{eq3}
\textbf{h}_{v,ST}^{t,l} = AGG_{across}(\{\textbf{h}_{v,R}^{t',l}|1 \leq t' \leq T\}; \Theta_{across})
\end{equation}
where $\textbf{h}_{v,R}^{t',l}$ is the spatial embedding of node $v$'s temporal neighbor at time period $t'$ in layer $l$, $\textbf{h}_{v,ST}^{t,l}$ is the spatial-temporal embedding of node $v$ at time period $t$ in layer $l$, $\Theta_{across}$ is the trainable parameters and is non-shareable for node type and aggregation layer, and $AGG_{across}(\cdot)$ is the aggregator function in across-time aggregation module.

\subsection{Prediction}
By stacking heterogeneous temporal aggregation layers $L$, we could derive the embeddings for developer nodes and file nodes in each time period, denoted $\textbf{h}_{v}^{t,L}$. We then simply sum the node embedding of all time periods as its final embedding:
\begin{equation}
\label{eq4}
\textbf{h}_{v} = \sum_{t=1}^T \textbf{h}_{v}^{t,L}
\end{equation}

Then the final embeddings of the developer nodes and the file nodes could be used at time period $T + 1$. In the test slice, the features of the developer nodes and the file nodes are the result of $L$ layers of spatial and temporal dependencies aggregation. 

During the prediction stage, the new issue nodes are processed by the spatial dependencies module, aggregating the feature information from developer-type and file-type neighbors to obtain the final features of the issue-type nodes. Finally, we calculate the inner product of the representations of issue node $v_{i}$ and developer node $v_{d}$ to predict their matching score as:
\begin{equation}
\label{eq5}
\hat{y}(v_{i},v_{d})=\textbf{h}_{v_{i}}^T \cdot \textbf{h}_{v_{d}}
\end{equation}

The automated issue assignment task is considered as a ranking problem and the loss function we adopt hinge ranking loss \cite{rennie2005fast}, which is more suitable for multi-label learning. This loss function is designed to maximize the difference between the positive and negative samples scores as much as possible.
\begin{equation}
\label{eq6}
L=\sum_{(v_{i},v_{d}^+,v_{d}^-) \in S} max(0, m-\hat{y}(v_{i},v_{d}^+)-\hat{y}(v_{i},v_{d}^-))
\end{equation}
where $m$ is a hyper-parameter that determines the margin, commonly set to 1.0 \cite{gong2013deep}. $\forall (v_{i},v_{d}^+) \in S^+$ denotes the set of observed interactions, that is, positive instances. $\forall (v_{i},v_{d}^-) \in S^-$ denotes the set of negative instances. $S=S^+ \cup S^-$.

\section{Experimental Setup}\label{experimental setup}
In this section, we present our experimental setup, mainly introducing the datasets, evaluation metrics, baselines, implementation details, and research questions.

\subsection{Datasets}

\begin{table*}[h]
\caption{Statistics of benchmark datasets.}
  \label{dataset}
\begin{tabular*}{\textwidth}{@{\extracolsep\fill}cccccccccc}
\toprule
Project & Time Range & $\#$issue & $\#$developer & $\#$file & $\# R_{report}$ & $\# R_{comment}$ & $\# R_{create}$ & $\# R_{remove}$ & $\# R_{similar}$ \\
\midrule
dotCMS\_core    & 03/2012-12/2018 & 13590 & 71 & 6830 & 23405 & 28712 & 1745 & 572 & 35678 \\
eclipse\_che        & 04/2015-11/2018 & 9756 & 129 & 12546 & 12163 & 24962 & 5075 & 2825 & 23276 \\
hazelcast\_hazelcast & 03/2012-12/2018 & 13131 & 198 & 10671 & 18376 & 39104 & 866 & 134 & 50604 \\
prestodb\_presto        & 08/2012-11/2018 & 10130 & 318 & 6948 & 12839 & 19485 & 2518 & 640 & 16038 \\
wildfly\_wildfly     & 09/2010-11/2018 & 11699 & 322 & 17202 & 21007 & 22444 & 10125 & 2088 & 18123 \\
\bottomrule
\end{tabular*}
\end{table*}

To evaluate our approach for automated issue assignment, we choose five GitHub repositories from the contribution database \cite{diamantopoulos2020employing} and relabel them as our benchmark datasets, which exhibit the highest number of issues within the contribution database.
Table \ref{dataset} shows the statistics of the datasets. The columns \textit{\#issue}, \textit{\#developer}, and \textit{\#file} show the number of each type of node for each project. 
The sixth to last columns depict the counts of the five relationships among the three types of nodes for each project.

\subsection{Evaluation Metrics} \label{evaluation metrics}
Referring to previous work \cite{mani2019deeptriage, jahanshahi2023adptriage, dong2024neighborhood, wang2024empirical}, we use the top-N hit rate and mean reciprocal rank (MRR) \cite{voorhees2001trec} as evaluation metrics.
Top-N hit rate calculates the correlation score between a given issue to be predicted and all candidate developers, and takes the top N developers as recommended fixers. In this paper, we select N in \{1, 3, 5\}.

MRR evaluates the effectiveness of the recommendation approach in accurately placing the first correct developer at the top of the recommended list, which is formulated as:
\begin{equation}
\label{Eq7}
MRR = \frac{1}{|I|} \sum_{i=1}^{|I|} \frac{1}{rank_i}
\end{equation}
where $rank_i$ denotes the position of the first correctly-recommended developer for the $i$-th issue, $|I|$ represents the total number of issue nodes.

Moreover, to assess the statistical significance of the performance differences between our approach and baselines, we employ the Wilcoxon signed-rank test \cite{wilcoxon1970critical}. $p$-value below 0.05 indicates that one approach significantly outperforms the other at a 95\% confidence level; otherwise, the difference is not considered statistically significant. Additionally, we use Cliff’s delta ($\delta$) \cite{macbeth2011cliff} to measure the effect size of the performance difference. The effectiveness level is categorized as negligible if $|\delta| < 0.147$, small if $0.147 \leq |\delta| < 0.33$, medium if $0.33 \leq |\delta| < 0.474$, and large if $|\delta| \geq 0.474$.

\subsection{Baselines} \label{baselines}
We select three types of baseline approaches in the mainstream for comparison, (1) content-based approaches: DeepTriage, Multi-triage; (2) expertise-based approach: ADPTriage; (3) CF-based approaches: PCG, NCGBT.

\begin{itemize}
\item{DeepTriage \cite{mani2019deeptriage}:} This approach uses an attention-based deep bidirectional recurrent neural network model to learn the syntactic and semantic features of issue reports in an unsupervised manner, and then employs a softmax layer to recommend potential developers as output.

\item{Multi-triage \cite{aung2022multi}:} This approach utilizes a multi-task learning framework to simultaneously resolve issue assignment and issue type allocation tasks. It incorporates a text encoder and an abstract syntax tree (AST) encoder to learn the precise representation of issue reports, thereby facilitating the recommendation of potential developers.

\item{ADPTriage \cite{jahanshahi2023adptriage}:} This approach enables real-time decision-making on issue assignments while taking into consideration developers' expertise, bug type, and bug-fixing time attributes without imposing any constraints on the stochastic process.

\item{PCG \cite{dai2024pcg}:} This approach utilizes graph collaborative filtering (GCF) \cite{he2020lightgcn} to model issue assignment task as predicting links in the bipartite graph of issue–developer correlations. In addition, it introduces prototype clustering-based augmentation to mitigate data sparsity and devise a semantic contrastive learning task to overcome semantic deficiency.

\item{NCGBT \cite{dong2024neighborhood}:} This approach models the relationships between issues and developers as a bipartite graph, and utilizes a pre-trained language model to acquire the initial representation of issue nodes. It employs a basic graph neural network (GNN) \cite{he2020lightgcn} framework to learn the representation of all nodes and leverage these representations to predict developers for a given issue.
\end{itemize}

\subsection{Implementation Details}
%All baselines and our proposed approach are implemented with Python 3.8.5 and PyTorch 2.1.0. All experiments are conducted on Windows Server 2016 Datacenter with 6 cores of 2.3 GHz CPU, 48 GB RAM and NVIDIA Tesla V100 GPUs with 32 GB memory.
We use the uniform distributed random initializer to generate the initial embeddings of the issue nodes, the developer nodes, and the file nodes.
We employ Adam optimizer \cite{kingma2014adam} with the learning rate set to 5e-3, and the weight decay set to 5e-4. For other parameters, we set the dropout rate to 0.2, the number of GNN layers to 2, and the hidden embedding dimension to 32. We also use ReLU as an activation function. We train our model with a fixed 100 epochs and use an early stopping strategy \cite{caruana2000overfitting} with a patience of 10. That is, the best models are selected when the validation loss does not decrease for 10 consecutive epochs.
We split the dataset into three sets with a ratio of 8:1:1. Specifically, we divide the issues into ten equal portions in chronological order, using the first eight portions for training, the second-to-last portion for validation, and the final portion for testing.
We frame our task to new issue-developer link prediction. The new issue-developer relation is defined as the issue-developer link that exists in time period $T + 1$ but not in time period $T$.
The size of the time window $tw$ is searched in \{1, 2, 3, 4, 5, 6, 7\}, which means, we consider the HTG of the past $tw$ time periods to predict the issue-developer relationship in the next time period. 

\subsection{Research Questions}
\begin{table*}[h]
\caption{Construct the different variants of our approach.}
  \label{variants}
\begin{tabular*}{\textwidth}{@{\extracolsep\fill}cl}
\toprule
%\multicolumn{2}{l}{The variants descriptions} \\
Variants & Descriptions \\
\midrule
$variant_0$ & Remove the issue-file and developer-file relationships, and only consider the issue-developer relationships ($R_{report}, R_{comment}$). \\
\midrule
$variant_1$ & Remove the issue-file relationships, and only consider the issue-developer relationships ($R_{report}, R_{comment}$) and developer-file \\ & relationships ($R_{create}, R_{remove}$). \\
\midrule
$variant_2$ & Remove the developer-file relationships, and only consider the issue-developer relationships ($R_{report}, R_{comment}$) and issue-file \\ & relationships ($R_{similar}$). \\
\midrule
\toolname & The model in this paper, considers the issue-developer relationships ($R_{report}, R_{comment}$), developer-file relationships \\ & ($R_{create}, R_{remove}$) and issue-file relationships ($R_{similar}$). \\
\bottomrule                
\end{tabular*}
\end{table*}

To evaluate the proposed approach, we study three research questions:

\textbf{RQ1: How effective is \toolname compared to baselines?}

In this RQ, we aim to give a global picture on the effectiveness of the proposed approach \toolname. We compare \toolname with five baseline approaches categorized into three types (i.e., content-based, expertise-based, and CF-based) listed in Section \ref{baselines}, and calculate the average top-N hit rate and MRR. To verify the significant difference in the comparison results, we use non-parametric statistical test methods Wilcoxon signed-rank test and Cliff’s $\delta$ (as introduced in Section \ref{evaluation metrics}) to assess the performance differences between \toolname and other baselines.
Notably, ADPTriage is evaluated solely using the top-N hit rate metric, as it differs from other baselines in that it does not rank candidate developers. Instead, it heuristically generates labels based on the expertise and availability of developers over time, without explicitly ordering the candidates, making the computation of the MRR metric infeasible.

\textbf{RQ2: How well does \toolname perform across developer groups?}

To answer RQ2, we first examine the issue resolution distribution in the test set and observe that a small group of highly active developers is responsible for a significant portion of issue resolutions. Based on this observation, we divide developers into two groups: core developers, who contribute more frequently, and non-core developers, who contribute less. We ensure that both groups are responsible for approximately the same number of resolved issues, allowing for a fair comparison of assignment accuracy across different groups of developer contribution.
We compare the performance of \toolname with the baselines for developers across the two groups. Notably, we did not compare with ADPTriage in this RQ. The primary reason is that ADPTriage employs a different labeling system compared to the other baselines. 
Unlike the traditional labeling system, ADPTriage does not consider the \textbf{\textit{assignee}} field of an issue as the final label for evaluation.  
This divergence in the label system makes a direct comparison with other approaches infeasible. However, we will discuss a potential comparison with ADPTriage in more detail in Section \ref{discussion}.

\textbf{RQ3: How do different types of relationships impact \toolname?} 

To answer RQ3, we design an ablation experiment to evaluate the impact of different types of relationships to \toolname. As detailed in Table \ref{variants}, we construct three additional variants: $variant_0$, which includes only the \textit{report} and \textit{comment} relationships between issues and developers; $variant_1$, which extends $variant_0$ by incorporating the \textit{create} and \textit{remove} relationships between developers and files; and $variant_2$, which builds on $variant_0$ by adding the \textit{similar} relationship between issues and files. \toolname combines all five relationships (\textit{report}, \textit{comment}, \textit{create}, \textit{remove}, and \textit{similar}) into a unified framework. 

\textbf{RQ4: What is the optimal time window size for \toolname?} 

\begin{figure}[!t]
\centering
\includegraphics[width=1.0\linewidth]{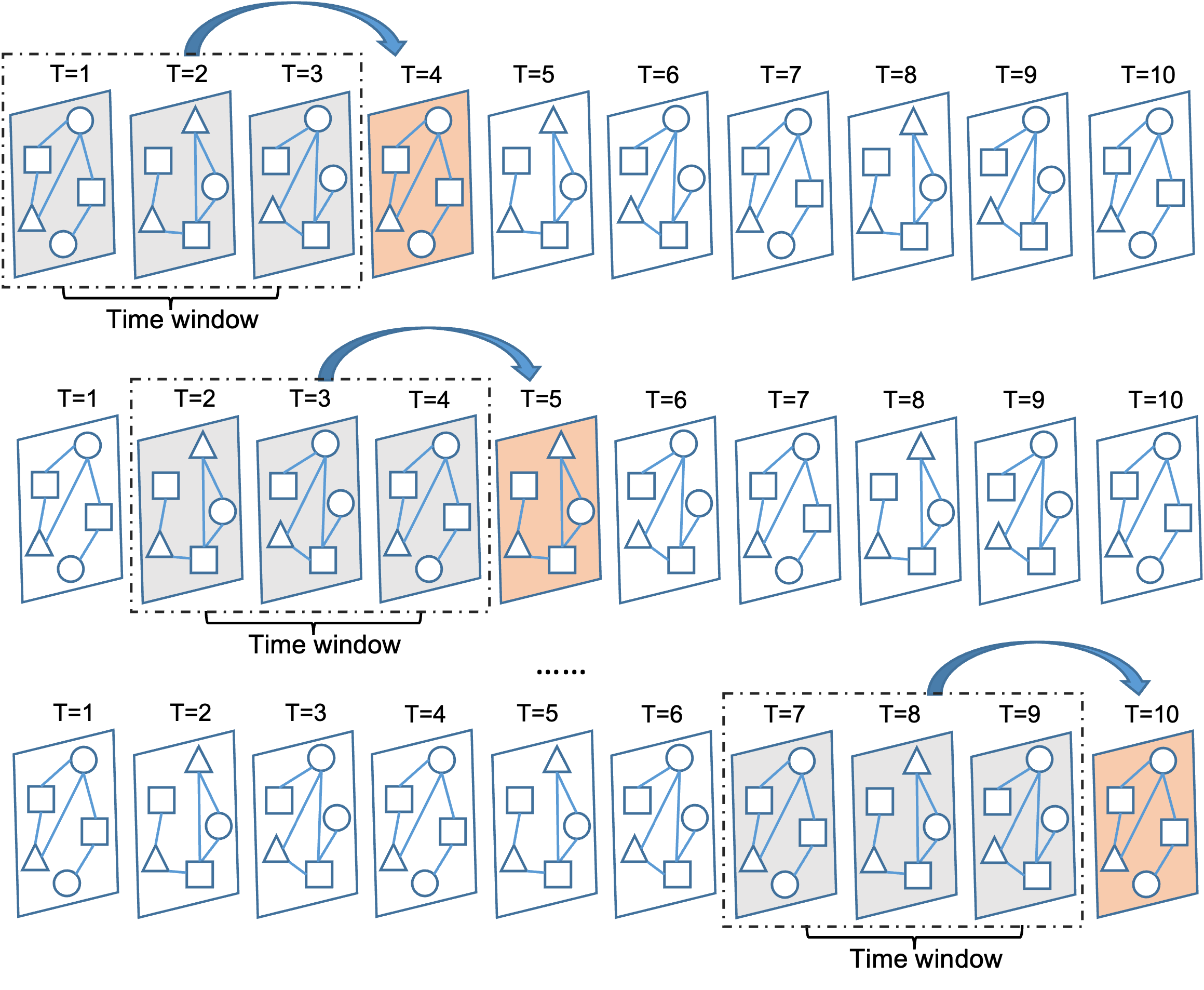}
\caption{The process of sliding time window.}
\label{RQ3-time-window}
\end{figure}

RQ4 aims to investigate how varying the size of the time window helps the model to better capture stage-specific patterns across different time stages. 
As illustrated in Fig. \ref{RQ3-time-window}, the dataset is divided into ten time slices. Assuming a time window size of 3, during training, the data within the time window is treated as known information to predict the issue-developer relationship in the subsequent time slice. For instance, time slices $(T1, T2, T3)$ are used as input to predict the issue-developer relation in $T4$, which constitutes one training batch. The time window then slides forward, using $(T2, T3, T4)$ to predict $T5$, and so on.
Since T9 and T10 are reserved as the validation and test sets, respectively, the maximum size of the time window is limited to 7. This ensures that there is at least one training batch, such as $(T1, T2, T3, T4, T5, T6, T7) \rightarrow T8$. This sliding time window approach enables the model to learn the data distribution at different time stages and dynamically capture changes in developer activity.

\section{Results and Analysis}\label{results and analysis}
\subsection{RQ1: How effective is \toolname compared to baselines?}

\begin{table*}[h]
\caption{Detailed comparison results of six approaches. The values in parentheses count the percentage of performance improvement of our approach over the best baseline.}
\label{RQ1-1}
\begin{tabular*}{\textwidth}{@{\extracolsep\fill}cccccc}
\toprule
\multirow{2}{*}{Project} & \multirow{2}{*}{Model} & \multicolumn{3}{c}{Top-N hit rate} & \multirow{2}{*}{MRR} \\\cmidrule{3-5} 
& & top-1 & top-3 & top-5 & \\
\midrule
\multirow{6}{*}{dotCMS\_core}
& DeepTriage & 0.195 & 0.409 & 0.523 & 0.113 \\
& Multi-triage & 0.167 & 0.306 & 0.418 & 0.138 \\
& ADPTriage & 0.349 & 0.604 & 0.720  & - \\
& PCG & 0.499 & 0.700 & 0.778 & 0.627 \\
& NCGBT & 0.507 & 0.658 & 0.777 & 0.624 \\
& \textbf{\toolname} & \textbf{0.576} $(\uparrow \textbf{13.64\%})$ & \textbf{0.797} $(\uparrow \textbf{13.88\%})$ & \textbf{0.828} $(\uparrow \textbf{6.43\%})$ & \textbf{0.700} $(\uparrow \textbf{11.76\%})$ \\
\midrule
\multirow{6}{*}{eclipse\_che} 
& DeepTriage & 0.437 & 0.599 & 0.686 & 0.185 \\
& Multi-triage & 0.469 & 0.609 & 0.672 & 0.194 \\
& ADPTriage & 0.143 & 0.300& 0.398 & - \\
& PCG & 0.552 & 0.658 & 0.694 & 0.624 \\
& NCGBT & 0.558 & 0.653 & 0.698 & 0.630 \\
& \textbf{\toolname} & \textbf{0.941} $(\uparrow \textbf{68.71\%})$ & \textbf{0.967} $(\uparrow \textbf{46.87\%})$ & \textbf{0.977} $(\uparrow \textbf{39.93\%})$ & \textbf{0.957} $(\uparrow \textbf{51.93\%})$ \\
\midrule
\multirow{6}{*}{hazelcast\_hazelcast} 
& DeepTriage & 0.265 & 0.437 & 0.534 & 0.330 \\
& Multi-triage & 0.225 & 0.383 & 0.469 & 0.212 \\
& ADPTriage & 0.233 & 0.320 & 0.451 & - \\
& PCG & 0.408 & 0.570 & 0.608 & 0.512 \\
& NCGBT & 0.446 & 0.593 & 0.651 & 0.540 \\
& \textbf{\toolname} & \textbf{0.847} $(\uparrow \textbf{90.05\%})$ & \textbf{0.916} $(\uparrow \textbf{54.34\%})$ & \textbf{0.930} $(\uparrow \textbf{42.86\%})$ & \textbf{0.886} $(\uparrow \textbf{64.22\%})$ \\
\midrule
\multirow{6}{*}{prestodb\_presto}    
& DeepTriage & 0.136 & 0.248 & 0.340 & 0.822\\
& Multi-triage & 0.141 & 0.267 & 0.361 & 0.461 \\
& ADPTriage & 0.483 & 0.483 & 0.483 & - \\
& PCG & 0.620 & 0.723 & 0.763 & 0.686 \\
& NCGBT & 0.651 & 0.735 & 0.773 & 0.712 \\
& \textbf{\toolname} & \textbf{0.844} $(\uparrow \textbf{29.74\%})$ & \textbf{0.877} $(\uparrow \textbf{19.19\%})$ & \textbf{0.896} $(\uparrow \textbf{15.96\%})$ & \textbf{0.869} $(\uparrow \textbf{5.63\%})$ \\
\midrule
\multirow{6}{*}{wildfly\_wildfly}   
& DeepTriage & 0.117 & 0.322 & 0.450 & 0.009 \\
& Multi-triage & 0.230 & 0.360 & 0.462 & 0.013 \\
& ADPTriage & 0.459 & 0.459 & 0.459 & - \\
& PCG & 0.495 & 0.599 & 0.625 & 0.560 \\
& NCGBT & 0.520 & 0.612 & 0.642 & 0.582 \\
& \textbf{\toolname} & \textbf{0.652} $(\uparrow \textbf{25.33\%})$ & \textbf{0.790} $(\uparrow \textbf{29.09\%})$ & \textbf{0.827} $(\uparrow \textbf{28.76\%})$ & \textbf{0.735} $(\uparrow \textbf{26.33\%})$ \\
\bottomrule
\end{tabular*}
\end{table*}

\begin{figure*}
\centering
\begin{minipage}[t]{1\textwidth}
\centering
    \includegraphics[width=0.49\textwidth]{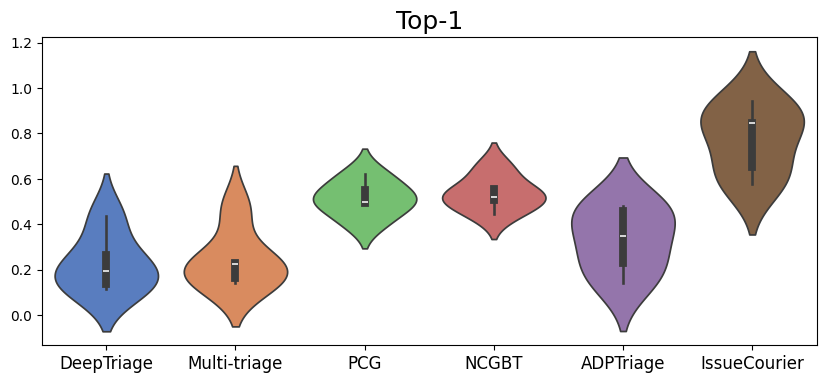}
    \includegraphics[width=0.49\textwidth]{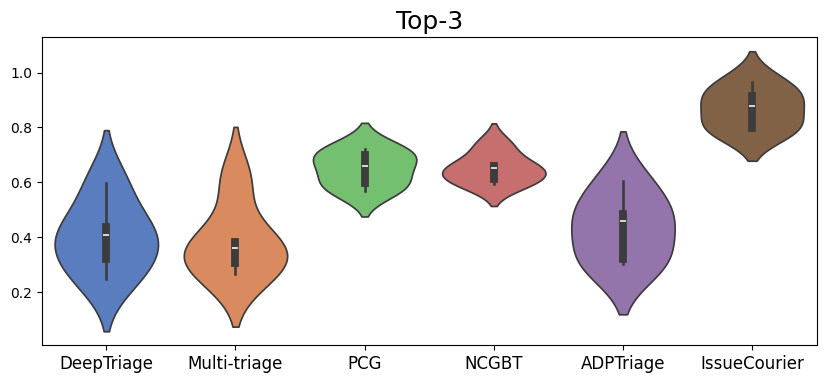}
\end{minipage}%

\begin{minipage}[t]{1\textwidth}
\centering
    \includegraphics[width=0.49\textwidth]{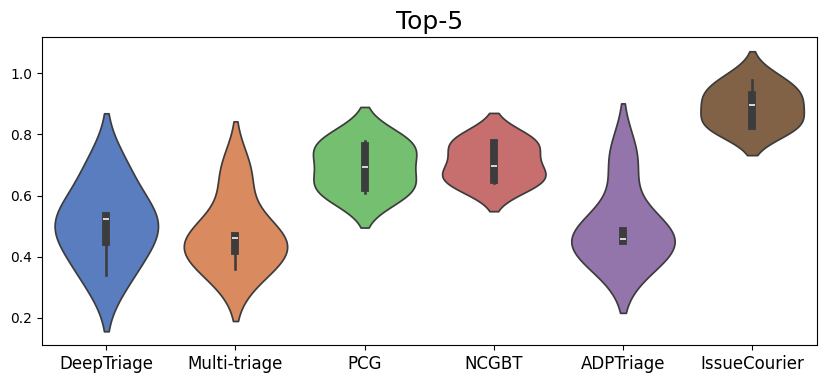}
    \includegraphics[width=0.49\textwidth]{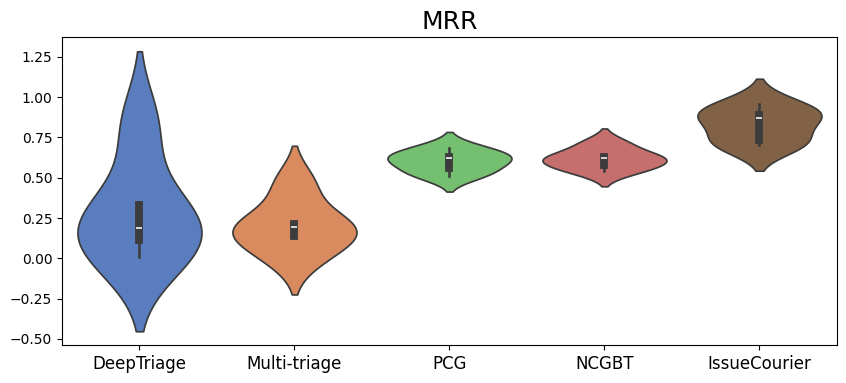}
\end{minipage}%

\caption{The top-N hit rate($N$ = 1, 3, 5) and MRR of six approaches over five projects.}
\label{RQ1-performance-fig}
\end{figure*}

\begin{table*}[h]
\caption{Statistical test results of our approach vs. baselines based on Wilcoxon signed-rank test and Cliff $\delta$, where avg denotes the average result of each approach on five projects. 
L denotes the large (i.e., $|\delta| \ge 0.474$) difference between our approach and baselines.
}
\label{RQ1-Wilcoxon-Cliff}
\begin{tabular*}{\textwidth}{@{\extracolsep\fill}ccccccccc}
\toprule
\multicolumn{1}{c}{Our approach} & \multicolumn{2}{c}{top-1} & \multicolumn{2}{c}{top-3} & \multicolumn{2}{c}{top-5} & \multicolumn{2}{c}{MRR}  \\ \cmidrule{2-3} \cmidrule{4-5} \cmidrule{6-7} \cmidrule{8-9}
\textit{vs} & avg & p-value ($\delta$) & avg & p-value ($\delta$) & avg & p-value ($\delta$) & avg & p-value ($\delta$)  \\
\midrule
DeepTriage & 0.230 & $\textbf{p\textless 0.05}$ (L) & 0.403 & $\textbf{p\textless 0.05}$ (L) & 0.507 & $\textbf{p\textless 0.05}$ (L) & 0.292 & $\textbf{p\textless 0.05}$ (L) \\
Multi-triage & 0.246 & $\textbf{p\textless 0.05}$ (L) & 0.385 & $\textbf{p\textless 0.05}$ (L) & 0.476 & $\textbf{p\textless 0.05}$ (L) & 0.204 & $\textbf{p\textless 0.05}$ (L) \\
ADPTriage & 0.333 & $\textbf{p\textless 0.05}$ (L) & 0.433 & $\textbf{p\textless 0.05}$ (L) & 0.502 & $\textbf{p\textless 0.05}$ (L) & - & - \\
PCG & 0.515 & $\textbf{p\textless 0.05}$ (L) & 0.650 & $\textbf{p\textless 0.05}$ (L) & 0.694 & $\textbf{p\textless 0.05}$ (L) & 0.602 & $\textbf{p\textless 0.05}$ (L)\\
NCGBT & 0.536 & $\textbf{p\textless 0.05}$ (L) & 0.650 & $\textbf{p\textless 0.05}$ (L) & 0.708 & $\textbf{p\textless 0.05}$ (L) & 0.618 & $\textbf{p\textless 0.05}$ (L)\\
\textbf{\toolname} & 0.772 & - & 0.869 & - & 0.892 & - & 0.830 & - \\
\bottomrule
\end{tabular*}
\end{table*}

To assess the effectiveness of \toolname, we conduct a comprehensive comparison against five issue assignment approaches, and the detailed results shown in Table \ref{RQ1-1}. The best performance of each project is highlighted in bold for clarity. 
Fig. \ref{RQ1-performance-fig} illustrates the distribution of top-N hit rate and MRR across the six approaches. The wider sections of the violin charts indicate a higher concentration of data in the area. The upper and lower bounds of the internal box within the violin chart represent the maximum and minimum values of top-N hit rate or MRR, while the central white short line represents the mean value.
Furthermore, we conduct the Wilcoxon signed-rank test and the effect size Cliff’s $\delta$, the results are shown in Table \ref{RQ1-Wilcoxon-Cliff}.

For the MRR score, \toolname achieves 0.700 to 0.957, outperforming all baselines. Compared to the best-performing baseline on the projects \textit{dotCMS\_core}, \textit{eclipse\_che}, \textit{hazelcast\_hazelcast}, \textit{prestodb\_presto} and \textit{wildfly\_wildfly}, the MRR of our approach improves by 11.76\%, 51.93\%, 64.22\%, 5.63\% and 26.33\%, respectively. 
For the top-N hit rate where $N$ is 1, 3, 5 in our experiments, we can find that \toolname achieves 0.576-0.941 in top-1, 0.790-0.967 in top-3, and 0.827-0.977 in top-5. Compared with the corresponding best baseline on each project, the top-1 of \toolname improves by 13.64\% on the \textit{dotCMS\_core} project, 68.71\% on \textit{eclipse\_che} project, 90.05\% on \textit{hazelcast\_hazelcast} project, 29.74\% on \textit{prestodb\_presto} project, and 25.33\% on \textit{wildfly\_wildfly} project, respectively. 
This indicates that \toolname can provide better recommendation results, especially when focusing on the top-ranked candidate.
As shown in Table \ref{RQ1-Wilcoxon-Cliff}, the statistical test results further confirm the superiority of \toolname. At a 95\% confidence level (i.e., p-value\textless 0.05) and an effect size of $|\delta| \ge 0.474$, \toolname consistently outperforms the baselines, demonstrating its robustness and effectiveness.

Next, we analyze how \toolname improves performance by addressing the limitations of existing approaches.

1) The content-based approaches, DeepTriage and Multi-Triage, do not achieve good performance in most cases. Compared with them, \toolname shows distinct improvements in terms of all metrics in all projects. This may be due to the fact that content-based approaches rely heavily on the quality of textual issue reports, making them susceptible to cases where the textual descriptions are missing or lack informative content.
In contrast, \toolname emphasizes modeling the interactions among issues, developers, and files through a HTG. By capturing these intricate relationships, \toolname effectively incorporates contextual and relational information overlooked by traditional text-focused approaches while mitigating the impact of low-quality issue reports, ultimately leading to significantly improved performance.

2) The CF-based approaches, PCG and NCGBT, achieve the best overall performance among the baselines. Their success can be attributed to the use of issue-developer relationships, which are modeled as a bipartite graph. By employing GNNs, these approaches effectively learn features from complex interactions between issues and developers. \toolname further extends this graph-based modeling by incorporating not only issue-developer relationships but also issue-file and developer-file interactions into a unified HTG. 
The HTG structure provides richer contextual information for developers and enables the model to learn their contribution patterns across different time periods. By capturing both spatial and temporal dependencies, \toolname alleviates data sparsity and enhances the model's ability to adapt to changes in developer activity over time.

3) Third, we observed that the expertise-based approach, ADPTriage, does not perform as well as PCG and NGCBT. However, it achieves better performance than DeepTriage and Multi-Triage in some cases.
Although ADPTriage considers developers' expertise, bug type, and developer availability, which expands the information to some extent, it still faces limitations due to the quality of issue reports. For a new issue with limited or uninformative textual content, ADPTriage tends to recommend experienced developers, without properly considering the developers who are directly relevant to the issue.
In contrast, \toolname incorporates multiple sources of information to more accurately recommend relevant developers. By associating developers with new issues, such as the reporter or commenters, and by leveraging developer familiarity with code modules related to the issue, \toolname is better equipped to recommend developers who are more relevant to the issue.  

\find{
% 45.49% = (13.64% + 68.71% + 90.05% + 29.74% + 25.33%) / 5
% 31.97% = (11.76% + 51.93% + 64.22% +5.63% + 26.33%) / 5
\textbf{RQ1 Summary:} \toolname surpasses the best-performing baseline by effectively leveraging multi-relational information and temporal slicing technique, which help extract richer contextual information and stage-specific developer activities to enhance recommendation accuracy.
}

\subsection{RQ2: How well does \toolname perform across developer groups?}

\begin{figure*}
\centering
\begin{minipage}[t]{0.40\linewidth}
\centering
    \includegraphics[width=\linewidth]{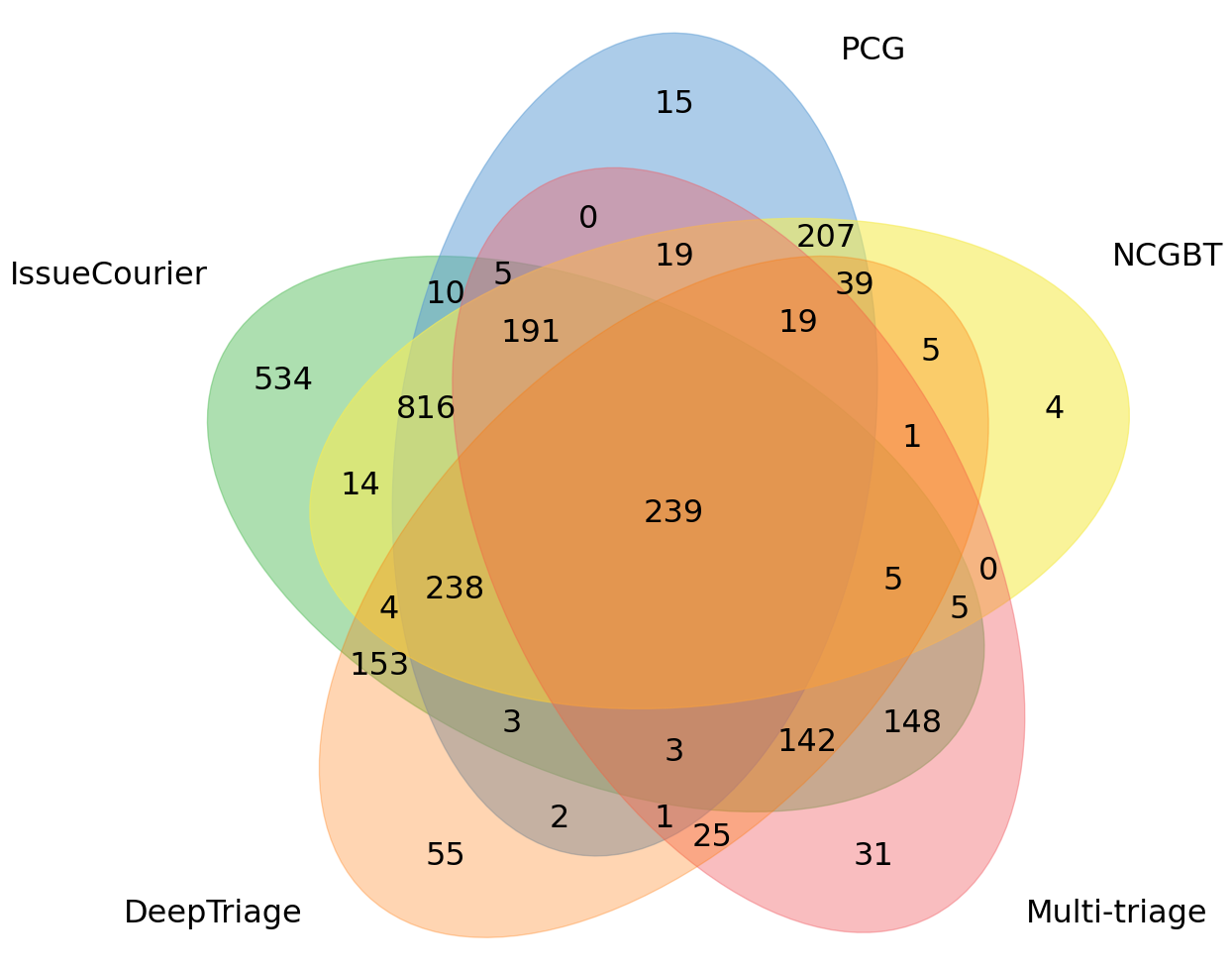}
    \caption*{(a) Core developer group} 
\end{minipage}%
\begin{minipage}[t]{0.40\linewidth}
\centering
    \includegraphics[width=\linewidth]{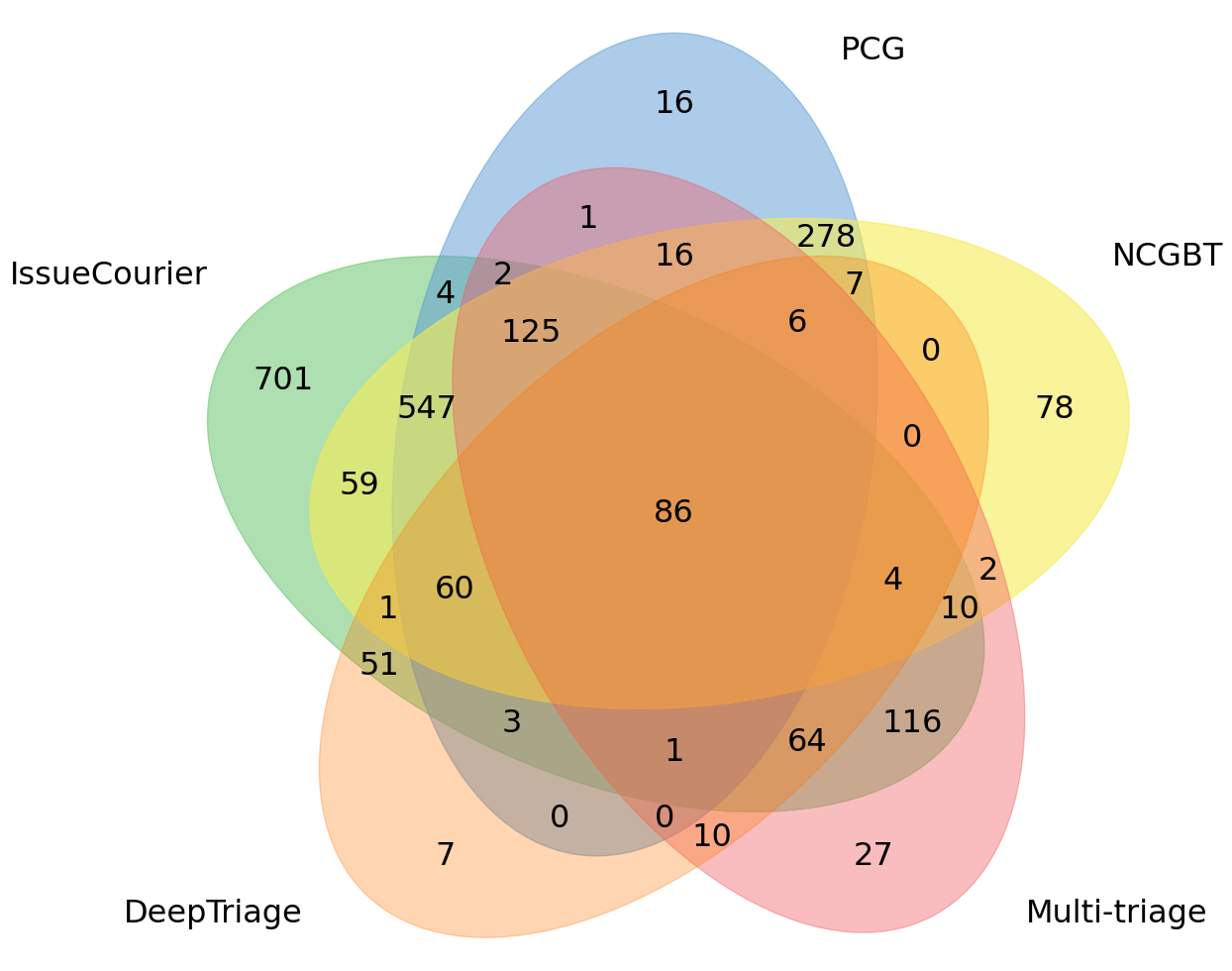}
    \caption*{(b) Non-core developer group} 
\end{minipage}%
\caption{Degree of orthogonality based on correct predictions of baselines across all dataset in term of different developer groups.}
\label{RQ2-venn}
\end{figure*}

\begin{table*}[!t]
\caption{Performance comparison of five approaches across different developer groups.}
\label{RQ2}
\begin{tabular*}{\linewidth}{@{\extracolsep\fill}cccccccc}
\toprule
Project & Model & $T_{core}$ & $T_{non-core}$ & $F_{core}$ & $F_{non-core}$ &  $Recall_{core}$ & $Recall_{non-core}$ \\
\midrule
\multirow{5}{*}{dotCMS\_core}
& DeepTriage   & 175  & 55 & 249 & 269 & 19.5\% & 7.7\% \\
& Multi-triage & 81 & 115 & 140 & 147 & 9.0\% & 16.1\% \\
& PCG          & 350 & 328 & 158 & 169 & 38.9\% & 45.8\% \\
& NCGBT        & \textbf{352} & 337 & 159 & 170 & \textbf{39.2\%} & 47.1\%  \\
& \textbf{\toolname} & 330 & \textbf{453} & \textbf{85} & \textbf{91} & 36.7\% $(\downarrow 6.3\%)$ & \textbf{63.3\%} $(\uparrow \textbf{34.4\%})$ \\
\midrule
\multirow{5}{*}{eclipse\_che} 
& DeepTriage   & 280 & 141 & 84 & 91 & 50.4\% & 25.5\% \\
& Multi-triage & 240 & 212 & 76 & 82 & 43.2\% & 38.3\%  \\
& PCG          & 314 & 224 & 51 & 56 & 56.5\% & 40.5\%  \\
& NCGBT        & 312 & 232 & 53 & 43 & 56.1\% & 42.0\% \\
& \textbf{\toolname} & \textbf{496} & \textbf{419} & \textbf{21} & \textbf{24} & \textbf{89.2\%} $(\uparrow \textbf{58.0\%})$ & \textbf{75.8\%} $(\uparrow \textbf{80.6\%})$ \\
\midrule
\multirow{5}{*}{hazelcast\_hazelcast} 
& DeepTriage   & 270 & 63 & 226 & 229 & 35.6\% & 10.2\% \\
& Multi-triage & 229 & 53 & 147 & 152 & 30.2\% & 8.6\% \\
& PCG          & 367 & 169 & 367 & 382 & 48.4\% & 27.4\% \\
& NCGBT        & 367 & 218 & 307 & 320 & 48.4\% & 35.4\% \\
& \textbf{\toolname} & \textbf{685} & \textbf{420} & \textbf{59} & \textbf{61} & \textbf{90.3\%} $(\uparrow \textbf{86.6\%})$ & \textbf{68.2\%} $(\uparrow \textbf{92.7\%})$ \\
\midrule
\multirow{5}{*}{prestodb\_presto}    
& DeepTriage   & 101 & 24 & 265 & 277 & 18.1.4\% & 4.3\% \\
& Multi-triage & 90  & 39 & 159 & 165 & 16.1\% & 7.0\% \\
& PCG          & 399 & 229 & 139 & 145 & 71.4\% & 41.3\% \\
& NCGBT        & 400 & 259 & 129 & 136 & 71.6\% & 46.8\% \\
& \textbf{\toolname} & \textbf{527} & \textbf{328} & \textbf{48}  & \textbf{87} & \textbf{94.3\%} $(\uparrow \textbf{31.8\%})$ & \textbf{59.2\%} $(\uparrow \textbf{26.6\%})$ \\
\midrule
\multirow{5}{*}{wildfly\_wildfly}   
& DeepTriage   & 108 & 17 & 417 & 418 & 16.5\% & 2.9\% \\
& Multi-triage & 194 & 51 & 220 & 220  & 30.0\% & 8.7\% \\
& PCG          & 377 & 202 & 321 & 322 & 57.6\% & 34.6\% \\
& NCGBT        & 375 & 233 & 286 & 287 & 57.3\% & 40.0\% \\
& \textbf{\toolname} & \textbf{472} & \textbf{290} & \textbf{137}  & \textbf{137} & \textbf{72.2\%} $(\uparrow \textbf{25.2\%})$ & \textbf{49.7\%} $(\uparrow \textbf{24.5\%})$ \\
\toprule
\multirow{5}{*}{SUM}   
& DeepTriage   & 934 & 300 & 1241 & 1284 & 27.3\% & 10.0\%\\
& Multi-triage & 834 & 470 & 742 &  766 & 24.3\% & 15.6\% \\
& PCG          & 1807 & 1152 & 1036 & 1074 & 52.7\% & 38.1\%\\
& NCGBT        & 1806 & 1279 & 934 & 956 & 52.7\% & 42.3\% \\
& \textbf{\toolname} & \textbf{2510} & \textbf{1910} & \textbf{350} & \textbf{400} & \textbf{73.2\%} $(\uparrow \textbf{38.9\%})$ & \textbf{63.2\%} $(\uparrow \textbf{49.3\%})$ \\
\bottomrule
\end{tabular*}
\begin{tablenotes}
\footnotesize
\item[a] \textbf{Note:} $T_{core}$ and $T_{non-core}$ represent the number of correctly predicted issues for the core and non-core developers, respectively. $F_{core}$ refers to issues that did not originally belong to the core developers but were incorrectly assigned to developers in this group. $F_{non-core}$ denotes the number of issues that should have been assigned to the non-core developers but were instead misclassified to the core developers. $Recall_{core}$ and $Recall_{non-core}$ measure the proportion of correctly assigned issues among all issues that actually belong to the core and non-core developers, respectively.
\end{tablenotes}
\end{table*}

To evaluate the effectiveness of \toolname across different developer groups, we compare it with four issue assignment approaches: DeepTriage, Multi-Triage, PCG, and NCGBT. The detailed results are presented in Table \ref{RQ2}. It is apparent that \toolname consistently achieves outstanding performance for both core and non-core developer groups.
Specifically, \toolname achieves 36.7\% to 94.3\% in $Recall_{core}$, and 59.2\% to 75.8\% in $Recall_{non-core}$. Compared to the best-performing baselines across the five datasets, \toolname improves $Recall_{core}$ by 25.2\% to 86.6\% (with the exception of the \textit{dotCMS\_core} project where PCG and NCGBT slightly outperform \toolname in the core group), and $Recall_{non-core}$ by 24.5\% to 92.7\%. In addition, $T_{core}$ and $T_{non-core}$ show that \toolname successfully recommends more correct assignments for both groups. From the overall results, \toolname achieves $Recall_{core}$ of 73.2\% and $Recall_{non-core}$ of 63.2\%, showing improvements of 38.9\% and 49.3\%, respectively, over the best-performing baseline. 

Fig. \ref{RQ2-venn} illustrates the results of the orthogonality analysis based on the correct predictions of baselines. The value in the common area indicates the cardinality of the set of issues from all datasets that are correctly assigned to developers by all approaches. 
The exclusive areas indicate the degree of orthogonality of each approach. 
In Fig. \ref{RQ2-venn} (a), we can find that all issues in the core developer group, 534 can be correctly recommended by \toolname only, while 55, 31, 15, 4 can be correctly recommended by DeepTriage, Multi-triage, PCG, and NCGBT only, respectively. A similar observation can be drawn from Fig. \ref{RQ2-venn} (b), of all issues in the non-core developer group, 701 can be correctly recommended by \toolname only, while 7, 27, 16, 78 can be correctly recommended by DeepTriage, Multi-triage, PCG, and NCGBT only, respectively. In other words, in addition to correctly predicting the same issues as the other baselines, \toolname successfully predicts many issues that the other baselines fail to assign correctly.

The results demonstrate that \toolname achieves more fair and accurate assignment performance for both core and non-core developers. In particular, \toolname effectively provides valuable information for developers with limited fixing histories, addressing a key limitation of traditional approaches, which struggle to capture useful insights about non-core developers.

\find{
\textbf{RQ2 Summary:} \toolname improves assignment accuracy across both developer groups, especially benefiting non-core developers, mitigating the impact of long-tailed data distribution and helping to provide more potential participation opportunities for non-core developers.
}

\subsection{RQ3: How do different types of relationships impact \toolname?}

In this experiment, we conduct a comprehensive ablation study to investigate the impact of different types of relationships on model performance. Here, we construct three additional variants, as shown in Table \ref{variants}. The comparison results are presented in Table \ref{RQ4}. It can be seen that \toolname demonstrates superior performance compared to the other three variants in most cases, achieving the highest average results for both the top-1 and MRR metrics. Specifically, the average top-1 and MRR of \toolname are 0.772 and 0.830, which are improved by 2.80\% and 1.34\% compared to the best-performing variant $variant_0$. 
$variant_1$ and $variant_2$, which incorporate additional relationships (e.g., developer-file and issue-file), perform worse than $variant_0$, which relies solely on the issue-developer relationships. This performance decline may be attributed to the auxiliary nature of the added relationships. Although issue-file and developer-file relationships provide potentially useful contextual information, they might introduce noise that undermines direct and critical information between issues and developers, leading to a decrease in predictive accuracy. When all relationships (issue-developer, issue-file, and developer-file) are integrated into \toolname, performance improves significantly. This enhancement could be due to the balanced and comprehensive integration of multiple relationships, which allows the model to extract complementary information while mitigating noise. 
It suggests that while individual auxiliary relationships may not always be beneficial in isolation, their combined use within a carefully designed framework can effectively enhance performance.

\find{
\textbf{RQ3 Summary:} By integrating multiple relationship types (issue-developer, issue-file, and developer-file), \toolname enriches the contextual information of nodes and enhances overall performance.
}

\begin{table}[!t]
\caption{Comparison of the impacts of different variants on performance.}
\label{RQ4}
\begin{tabular*}{\linewidth}{@{\extracolsep\fill}ccccc}
\toprule
\multirow{2}{*}{Variants} & \multicolumn{3}{c}{Top-N hit rate} & \multirow{2}{*}{MRR} \\\cmidrule{2-4}  & top-1 & top-3 & top-5 & \\
\midrule
$variant_0$ & 0.751 & \textbf{0.869} & \textbf{0.895} & 0.819 \\
$variant_1$ & 0.737 & 0.835 & 0.870 & 0.799 \\
$variant_2$ & 0.742 & 0.840 & 0.871 & 0.804 \\
\textbf{\toolname} & \textbf{0.772} & \textbf{0.869} & 0.892 & \textbf{0.830} \\
\bottomrule
\end{tabular*}
\end{table}

\subsection{RQ4: What is the optimal time window size for \toolname?}

\begin{figure}[!t]
\centering
\includegraphics[width=1.0\linewidth]{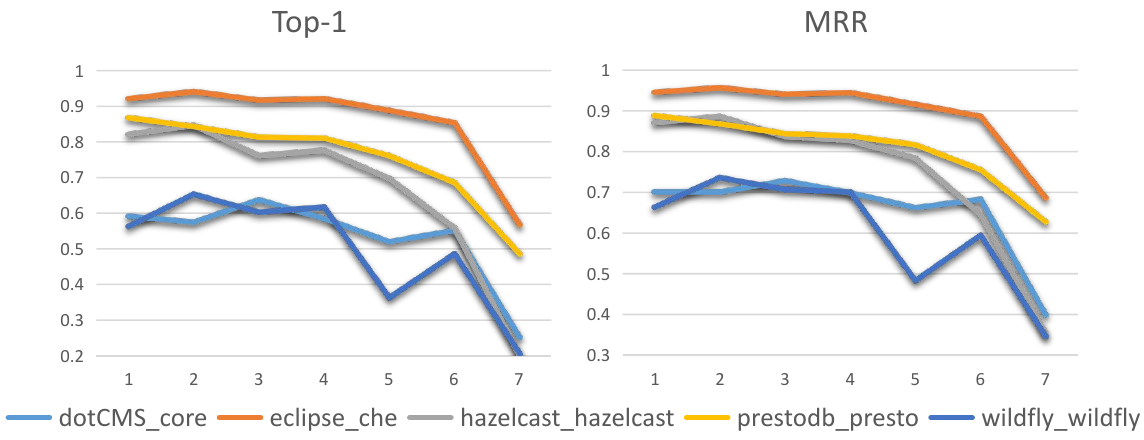}
\caption{Performance comparison over different time window size.}
\label{RQ3-fig}
\end{figure}

Table \ref{RQ3} shows the overall performance of the five projects in varying time window sizes, and Fig. \ref{RQ3-fig} presents the performance trends of each project in different time window sizes. We have the following observations. It can be observed that when the time window size $tw$ ranges from 1 to 4, the performance of the model shows minimal differences, the optimal setting differing between projects. For example, $tw=1$ performs best for \textit{prestodb\_presto} project, $tw=2$ is optimal for \textit{eclipse\_che}, \textit{hazelcast\_hazelcast}, and \textit{wildfly\_wildfly} projects, and $tw=3$ works best for \textit{dotCMS\_core} project. Statistical test results further confirm that there is no significant performance difference between $tw=1$ and $tw=2$ ($p$-value = 0.4114, $\geq 0.05$), between $tw=2$ and $tw=3$ ($p$-value = 0.3598, $\geq 0.05$), and between $tw=3$ and $tw=4$ ($p$-value = 0.7508, $\geq 0.05$). On average, $tw=2$ proves to be the most effective, achieving 0.772 in top-1 and 0.830 in MRR. As the time window size exceeds 4, the performance of the model declines significantly, with the $p$-value = 0.0479 ($\leq 0.05$) between $tw=2$ and $tw=5$ in statistical tests. This decline underscores the adverse impact of inactive developers on predictions as the time window expands, highlighting the importance of selecting an appropriate time window size to achieve a balance between the influence of developers' early contributions and recent activities.

\find{
\textbf{RQ4 Summary:} \toolname achieves its best performance with a time window size of 2, which effectively balances the influence of developers' recent activity and historical contributions, while performance drops notably when the window size exceeds this threshold.
}

\begin{table}[!t]
\caption{Effect of time window size.}
\label{RQ3}
\begin{tabular*}{\linewidth}{@{\extracolsep\fill}ccccc}
\toprule
\multirow{2}{*}{Time Window} & \multicolumn{3}{c}{Top-N hit rate} & \multirow{2}{*}{MRR} \\\cmidrule{2-4}  & top-1 & top-3 & top-5 & \\
\midrule
1 & 0.753 & 0.850 & 0.876 & 0.814 \\
2 & \textbf{0.772} & \textbf{0.869} & \textbf{0.892} & \textbf{0.830} \\
3 & 0.747 & 0.863 & 0.888 & 0.812 \\
4 & 0.743 & 0.842 & 0.865 & 0.802 \\
5 & 0.647 & 0.786 & 0.832 & 0.733 \\
6 & 0.628 & 0.771 & 0.813 & 0.714 \\
7 & 0.344 & 0.544 & 0.660 & 0.484 \\
\bottomrule
\end{tabular*}
\end{table}

\section{Discussion}\label{discussion}

In this section, we compare \toolname with ADPTriage under different labeling systems.
Traditional approaches determine the ground truth developer based on the issue's \textit{assignee} field. Our work relabels the ground truth by identifying the actual fixer of the issue. 
The traditional labeling system is precise and deterministic, as the developer who fixes the issue is labeled as the ground truth.
However, ADPTriage does not adhere to a deterministic labeling system. Instead, its labels are generated heuristically based on the expertise and availability of developers over time, meaning that the ground-truth developer for an issue remains uncertain until ADPTriage makes a decision.
This results in a more flexible but also less deterministic labeling approach, introducing a degree of randomness in the assignment process. 

In our previous RQ1 analysis, ADPTriage was evaluated using its own heuristic labeling system. 
In this section, we aim to further explore ADPTriage’s performance under traditional deterministic label system to assess its effectiveness in a more conventional evaluation setting. 
Table \ref{discussion} presents the results of ADPTriage when evaluated using the traditional labeling system. It can be observed that ADPTriage correctly predicted a total of 354 issues across the five projects with top-5 hits, and the highest top-5 hit rate reached only 0.1558 for the \textit{eclipse\_che} project. These results indicate a significant discrepancy between ADPTriage’s heuristic labeling system and the conventional ground truth system. Although ADPTriage’s labeling system introduces flexibility by considering developers based on their experience and availability, our findings suggest that these heuristic assignments often do not align with the actual fixers.

One possible reason for this gap lies in the underlying relationships between developers and issues. Although historical expertise provides useful information on the ability of a developer to address certain issues, additional contextual factors must be considered. Specifically, direct interactions between developers and issues, such as whether a developer has reported or commented on an issue, can be a stronger indicator of the likelihood of assignment. Moreover, the collaboration patterns among developers, such as how frequently certain developers co-comment on the same issues, may also play a crucial role in determining the most suitable assignee. These observations suggest that an effective issue assignment model should not only rely on historical expertise but also take into account the structural relationships between developers and issues. By evaluating ADPTriage under traditional labeling system, we gain a clearer understanding of how different labeling frameworks impact performance evaluation. Although heuristic approaches offer flexibility, they may introduce inconsistencies compared to deterministic ground-truth labels. 

\begin{table}[!t]
\caption{Performance of ADPTriage under the traditional labeling system.}
\label{discussion}
\begin{tabular*}{\linewidth}{@{\extracolsep\fill}ccccc}
\toprule
\multirow{2}{*}{Project} & \multicolumn{3}{c}{Top-N hit rate} & \multirow{2}{*}{Top-5 hits} \\\cmidrule{2-4}  & top-1 & top-3 & top-5 & \\
\midrule
dotCMS\_core & 0.0097 & 0.0507 & 0.0918 & 114 \\
eclipse\_che & 0 & 0.1536 & 0.1558 & 143 \\
hazelcast\_hazelcast & 0 & 0.0126 & 0.0763 & 97 \\
prestodb\_presto & 0 & 0 & 0 & 0 \\
wildfly\_wildfly & 0 & 0 & 0 & 0 \\
\bottomrule
\end{tabular*}
\end{table}

\section{Threats to Validity}\label{threats to validity}
The first threat to validity lies in the manual assessment of our relabeled datasets. To mitigate this threat, one of the authors of this paper and a graduate student independently labeled a randomly sampled subset of issues and achieved a Cohen’s Kappa score of 0.89. Both evaluators have 3 to 5 years of programming experience, providing sufficient expertise to conduct reliable labeling. Ultimately, they discussed the inconsistencies between their inspections and reached a uniform result.

The second threat to validity comes from the implementations of baselines. Many prior studies utilize different datasets (and not GitHub) and are optimized under different scopes, making direct comparisons infeasible. To address this, we reproduced the baseline approaches using our dataset to support a fair and comparable evaluation. During this process, we followed the guidelines of the original papers and the parameter settings as closely as possible to minimize potential inconsistencies arising from re-implementation.

The last threat to validity is about the generalizability of our approach. All projects in our evaluation are open-source repositories written in Java and hosted on GitHub. As a result, our findings may not be generalized to other programming languages, issue trackers, and closed-source software.
Nevertheless, our approach is language-agnostic design, in which mathematical modeling and feature extraction are independent of the programming language. To further validate the generalizability of our approach, future work should explore its application to proprietary projects and repositories written in other programming languages.

\section{Related Work}\label{related work}

\textbf{Content-based.} Content-based approaches widely consider bug triaging as a classification problem, utilizing machine learning (ML) algorithms such as Naive Bayes (NB) and Support Vector Machines (SVM) with TF-IDF, to treat issue title and description as features \cite{murphy2004automatic, anvik2006should, anvik2011reducing, oliveira2021issue}.
In an early work, Murphy et al. \cite{murphy2004automatic} represented bug reports using a bag-of-words model and trained a NB classifier to recommend candidate developers for new bugs.
Later, Anvik et al. \cite{anvik2006should, anvik2011reducing} extended the work of Murphy et al. \cite{murphy2004automatic} by testing six different ML algorithms for bug assignment. The results indicated that both the NB and SVM algorithms achieved the highest precision in a single recommendation scenario, while SVM performed better when generating multiple recommendations. 
With the development of deep learning, various classification techniques have been introduced to enhance issue assignment \cite{lee2017applying, mani2019deeptriage, he2021automatic, choquette2019multi, aung2022multi, lee2022light}.
For example, Mani et al. \cite{mani2019deeptriage} introduced a deep bidirectional recurrent neural network with an attention mechanism (DBRNN-A) to improve bug triaging by capturing syntactic and semantic relationships in bug reports, which utilized Word2Vec \cite{mikolov2013efficient} for text vectorization and training classifiers on the extracted features.
Aung et al. \cite{aung2022multi} introduced a multi-triage model based on multi-task learning, which simultaneously assigned developers and classified issue types, using a CNN-based text encoder and a BiLSTM-based AST encoder for feature extraction, and a contextual data augmentation approach to balance class distributions.
However, content-based approaches relied heavily on the quality of issue report text, making them less effective when reports are incomplete, vague or lack sufficient contextual details to accurately recommend the appropriate developer.

\textbf{CF-based.} To address the challenges of insufficient issue report quality, some researchers leveraged the interaction between issues and developers, as developer expertise is reflected in their issue-fixing records \cite{dai2024pcg}. Inspired by collaborative filtering-based information retrieval \cite{he2017neural, zhu2024representation}, studies have explored representing developers by mining issue-developer interaction relationships \cite{jeong2009improving, park2011costriage, zhang2016ksap, fang2021effective, jahanshahi2022s, dai2023graph, dai2024pcg, dong2024neighborhood}.
Park et al. \cite{park2011costriage} introduced CosTriage, an early attempt to model bug-developer correlations as links on a graph, using topic words extracted from bug reports to construct a tripartite graph of bug, topic, and developer. 
Jahanshahi et al. \cite{jahanshahi2022s} proposed the Schedule and Dependency-aware Bug Triage (S-DABT) method, integrating integer programming and ML to assign bugs to developers while considering textual data, bug dependencies, and developer schedules, improving bug resolution times and reducing overdue bugs.
Dai et al. \cite{dai2023graph} introduced the Graph Collaborative Filtering based Bug Triaging (GCBT) framework, which models bug-developer relationships as a bipartite graph, using a GRU-based NLP module and a spatial-temporal graph convolution strategy for improved bug triaging. In a subsequent work, Dai et al. \cite{dai2024pcg} proposed the PCG framework, enhancing bug-developer correlations by integrating prototype augmentation and contrastive learning into GCF to address data sparsity and semantic deficiency.
Dong et al. \cite{dong2024neighborhood} introduced the Neighborhood Contrastive Learning-based Graph Neural Network Bug Triaging (NCGBT) framework, which incorporated neighborhood contrastive learning from both structural and semantic perspectives to improve bug-developer relationship modeling, further enhancing bug triaging.
However, a common limitation of CF-based approaches is their reliance on explicit historical interactions between developers and issues, which can lead to performance degradation when such interactions are sparse or insufficient, limiting their ability to make accurate predictions for less active developers.

\textbf{Expertise-based.} 
Several studies have explored expertise-based approaches, which prioritize assigning issues to developers based on their skills, experience, and historical performance, helping identify the suitable developer for the task based on their expertise \cite{naguib2013bug, gousios2014exploratory, wang2017recommending, amreen2019developer, yadav2019ranking, sajedi2020vocabulary, da2020developer, jahanshahi2023adptriage, yadav2024developer}.
Gousios et al. \cite{gousios2014exploratory} analyzed various GitHub projects and found that factors such as developers' expertise, reputation, and contributions influenced pull request merge decisions, highlighting the importance of expertise in task allocation.
Amreen et al. \cite{amreen2019developer} developed a reputation-based developer recommendation approach that integrated data from public Git repositories, consolidating developer profiles by resolving identity discrepancies and measuring the impact of their work to create reputation badges, thereby enhancing trust and prioritizing pull requests and commits.
Jahanshahi et al. \cite{jahanshahi2023adptriage} developed ADPTriage, an ADP-based bug triage solution that optimized real-time bug assignments considering developers' expertise, bug types, and fixing times while accounting for uncertainties in bug arrivals and developers' schedules, aiming to support efficient decision-making in dynamic open-source projects.
Yadav et al. \cite{yadav2024developer} introduced a work engagement-sensitive bug triage approach (DLB) that assigns bugs to developers based on technical skills, work engagement, and current workload to improve triage efficiency while balancing developer workload.
However, a common limitation of expertise-based approaches is their tendency to recommend experienced developers, often overlooking less active or newer contributors. This bias results in an uneven distribution of issue assignments, where a small group of experienced developers receive the majority of recommendations, while more than 80\% developers have minimal opportunities to be recommended \cite{asthana2019whodo}.

\section{Conclusion}\label{conclusion}

In this paper, we propose \toolname, a novel multi-relational embedded approach for automatic issue assignment, and provide a benchmark dataset which addresses the problem of incorrect and missing labels in existing OSS projects. 
\toolname formally defines the \textit{report} and \textit{comment} relationships between developers and issues, the \textit{create} and \textit{remove} relationships between developers and source code files, and the \textit{similar} relationship between issues and source code files, and explicitly encodes them into the representation learnings of issues and developers. Moreover, \toolname leverages a temporal slicing technique to divide the constructed heterogeneous graph into multiple time slices, capturing stage-specific patterns across different time periods. To evaluate the benefits of \toolname, we conduct extensive experiments on the benchmark dataset, demonstrating that \toolname significantly outperforms the state-of-the-art baselines. Our results show that \toolname improves top-1 and MRR by up to 45.49\% and 31.97\%, respectively, over the best-performing baseline.
Future work will explore refining the temporal modeling strategy, incorporating additional contextual information, and extending our model to more diverse OSS projects and proprietary projects. 

% \section*{Acknowledgments}
\section*{Acknowledgments}
This work was partially supported by National Key R\&D Plan of China (Grant No.2024YFF0908003), Natural Science Foundation of China (No.62472326), and CCF-Zhipu Large Model Innovation Fund (No.CCF-Zhipu202408).

\bibliographystyle{IEEEtran}
\balance
\bibliography{mybibfile}% common bib file

\end{document}